\documentclass[11pt]{article}

\usepackage{fullpage, amsfonts, graphicx, dsfont}
\usepackage{authblk}
\usepackage{algorithm}
\usepackage{algorithmic}
\usepackage[title]{appendix}

\title{Automated Design of Pulse Sequences for Magnetic Resonance Fingerprinting using Physics-Inspired Optimization}

\author[1]{Stephen~P.~Jordan}
\author[2]{Siyuan~Hu}
\author[3]{Ignacio~Rozada}
\author[2]{Debra~F.~McGivney}
\author[4]{Rasim~Boyacio{\u g}lu}
\author[5]{Darryl~C.~Jacob}
\author[2]{Sherry~Huang}
\author[1]{Michael~Beverland}
\author[6,1]{Helmut~G.~Katzgraber\thanks{The work of H.~G.~K. was performed before joining Amazon Web Services.}}
\author[1]{Matthias~Troyer}
\author[4]{Mark~A.~Griswold}
\author[2,7]{Dan Ma}
\affil[1]{\small Microsoft Quantum, Redmond WA}
\affil[2]{\small Biomedical Engineering, Case Western Reserve University, Cleveland OH}
\affil[3]{\small 1QBit, Vancouver BC}
\affil[4]{\small Radiology Department, Case Western Reserve University, Cleveland OH}
\affil[5]{\small Texas A \& M University, College Station TX}
\affil[6]{\small Professional Services, Amazon Web Services, Seattle WA}
\affil[7]{\small To whom correspondence should be addressed: \texttt{dan.ma@case.edu}}

\date{Physical Sciences: Engineering. Keywords: Magnetic Resonance Imaging, optimization, pulse sequence design, magnetic resonance fingerprinting}
\begin{document}

\bibliographystyle{unsrt}

\newcommand{\captionfonts}{\small}                    
\renewcommand{\th}{^\mathrm{th}}                      
\newcommand{\eq}[1]{(\ref{#1})}                       
\newcommand{\sect}[1]{\S\ref{#1}}                     
\newcommand{\ket}[1]{\left| #1\right\rangle}          
\newcommand{\bra}[1]{\left\langle #1\right|}          
\newcommand{\id}{\mathds{1}}                          
\newcommand{\tri}{s}                                  

\maketitle

\begin{abstract}
Magnetic Resonance Fingerprinting (MRF) is a method to extract quantitative tissue properties such as $T_1$ and $T_2$ relaxation rates from arbitrary pulse sequences using conventional magnetic resonance imaging hardware. MRF pulse sequences have thousands of tunable parameters which can be chosen to maximize precision and minimize scan time. Here we perform de novo automated design of MRF pulse sequences by applying physics-inspired optimization heuristics. Our experimental data suggests systematic errors dominate over random errors in MRF scans under clinically-relevant conditions of high undersampling. Thus, in contrast to prior optimization efforts, which focused on statistical error models, we use a cost function based on explicit first-principles simulation of systematic errors arising from Fourier undersampling and phase variation. The resulting pulse sequences display features qualitatively different from previously used MRF pulse sequences and achieve fourfold shorter scan time than prior human-designed sequences of equivalent precision in $T_1$ and $T_2$. Furthermore, the optimization algorithm has discovered the existence of MRF pulse sequences with intrinsic robustness against shading artifacts due to phase variation.
\end{abstract}

\section{Significance}

Magnetic resonance is a widely used non-invasive medical imaging technology. Most clinical magnetic resonance imaging scans generate qualitative or `weighted' images. A recent new technology, Magnetic Resonance Fingerprinting \cite{R1}, simultaneously extracts quantitative voxel-by-voxel measurements of multiple intrinsic tissue properties such as $T_1$ and $T_2$ relaxation rates in a single scan, rapid enough for clinical use. These have been used for tumor characterization \cite{BYD17, YBP17, HDG20}, epilepsy lesion detection \cite{MJD18,LWC18}, and for estimating variability of tissues in asymptomatic subjects \cite{BYR15, FBC20,WBG18}. Here, we find that detailed computer models of random and systematic errors can be combined with physics-inspired optimization heuristics to discover novel MRF pulse sequences achieving substantially improved performance relative to MRF pulse sequences previously designed by human experts.

\section{Introduction}

In contrast to traditional magnetic resonance imaging, which relies on simple pulse sequences with analytically solvable dynamics, Magnetic Resonance Fingerprinting extracts tissue properties such as $T_1$ and $T_2$ from arbitrary pulse sequences by pattern matching measured signals to a dictionary of numerically computed signals for different tissue types. This can be done by interpreting the signals as complex vectors, and finding the dictionary entry whose inner product with the observed signal has the largest magnitude \cite{R1}. By allowing arbitrary pulse sequences, MRF opens up a design space of thousands of tunable parameters over which to search for pulse sequences achieving greater precision at shorter scan time.

Optimization of MRF acquisition parameters, such as radio frequency pulses, timing, and magnetic field gradients, is needed to achieve the best signal to noise ratio, image quality, precision, and reproducibility. However, the design and optimization of pulse sequences is very challenging for two main reasons. First, it is difficult to design an efficiently computable cost function that accurately predicts in vivo performance of MRF pulse sequences. Second, the resulting optimization problem is computationally difficult, as the space of possible pulse sequences is too high-dimensional and non-convex for simple methods such as exhaustive search or gradient descent. In this work we use optimization algorithms to choose the flip angle and TR duration for each TR. Thus, the search space for a pulse sequence of $n$ pulses is $2n$-dimensional, where for 2D scans $n$ is typically on the order of 1000, and is even larger for 3D scans. 

In clinical settings it is typical to accelerate MRF scans by sampling only a small fraction of the relevant Fourier coefficients (``$k$-space'') after each pulse. For example, acceleration factors of 48 to 400 (sampling only 2\% to 0.2\% as compared to the Nyquist sampling requirement) have been reported in 2D or 3D MRF in vivo scans \cite{JMS15, MPJ16, MJC18, MJD18}. Through extensive in vivo experimentation, we find that the dominant sources of error in MRF brain scans in this regime are Fourier undersampling artifacts and system-based phase variation induced shading artifacts. The interplay of these two errors results in temporally and spatially dependent artifacts in the reconstructed images and causes aliasing and shading artifacts in the resulting tissue maps \cite{MJD18}. Random error due to background noise in the receive coils is also present but appears to play a secondary role.

To design MRF pulse sequences through optimization one first needs a cost function, which, given a proposed pulse sequence, produces a metric of its predicted effectiveness. In previous work, several cost functions have been proposed. Under the assumption of zero-mean independent Gaussian random error at each timestep in the raw signal, lower bounds on the standard deviation in the inferred values of $T_1$ and $T_2$ from an MRF scan can be obtained using the Cramer-Rao bound. Taking a linear combination of these bounds as a cost function, optimized MRF pulse sequences are obtained in \cite{R4} by sequential quadratic programming, in \cite{R7} using the BFGS algorithm, in \cite{STH17} using an interior point method, and in \cite{MPA16} using dynamic programming. In \cite{R8} the Gaussian model of random error is supplemented by a model of Fourier undersampling artifacts in which an additional Gaussian noise term is added whose magnitude is signal-dependent. Optimized sequences are then obtained using a genetic algorithm. In \cite{R5} the magnetization vs pulse index time-series for a given tissue is interpreted as a vector and inner product between the normalized vectors determined by different tissues is interpreted as a metric of distinguishability. Minimization of these inner products is then performed with simulated annealing, branch and bound, an interior point method, and brute force search. Several works have been devoted exclusively to the design of a cost function predictive of in vivo performance of MRF sequences, without also pursuing optimization \cite{R9, R10, R11}.

In the highly undersampled regime most relevant to clinical settings, we find that cost functions based on simple statistical noise models or crude heuristics such as minimizing inner product between signal vectors are insufficient to accurately predict in vivo performance of MRF pulse sequences. An obvious alternative is to use a comprehensive first-principles computer simulation of the MRF scan using explicit models of a tissue distribution, a choice of acquisition parameters including $k$-space trajectories, and an image reconstruction algorithm. Such computer models have been constructed in prior work and are referred to as ``digital phantoms\footnote{The name ``digital phantom'' is an analogy with traditional magnetic resonance phantoms, which are precisely-characterized physical artifacts used to calibrate MRI machines.}.'' Digital phantoms that directly model undersampling artifacts by carrying out non-uniform Fourier transforms are computationally costly, often taking several minutes to evaluate even on powerful workstations. This makes them challenging to use as a cost function within an optimization, as our computational experiments show that even carefully tuned optimization algorithms often require tens of thousands of cost function evaluations to find good pulse sequences.

Here we introduce an accelerated digital phantom which makes use of a simplified model of brain tissue distribution. This approximation speeds up the evaluation by approximately two orders of magnitude. Further detail on on the accelerated digital phantom is given in the Methods section and supporting appendices. A preliminary report also appears in \cite{R12}.

Our computational experiments show that, even when allowed thousands of cost function evaluations, standard ``off the shelf'' optimization methods such as sequential quadratic programming and BFGS yield poor pulse sequences. However, with careful parameterization of the search space, judicious generation of moves within the search space, and well-tuned hyperparameters, we find that good pulse sequences can be found with both simulated annealing and substochastic Monte Carlo optimization methods. Use of Monte Carlo methods, such as simulated annealing, for optimizing magnetic resonance protocols has a long history \cite{HBO69}. Our experiments show that such methods can be used to design novel high-speed magnetic resonance fingerprinting pulse sequences ``from scratch,'' \emph{i.e.} without human-designed pulse sequences as starting points, but that this requires careful design of the cost function as well as the moves by which the search space is explored, as discussed in the Methods section. In recent work, reinforcement learning has been used for de novo design of MRI pulse sequences \cite{ZLK18}.

Using these methods, the optimization algorithms yield pulse sequences that display superior performance to prior state of the art MRF pulse sequences designed by human experts. Improved precision can be achieved by increasing the number of pulses (and hence the scan time). Thus, a performance comparison can be made by plotting optimized pulse sequences against the precision vs duration tradeoff curve obtained by taking truncations of the human-designed pulse sequence to different numbers of pulses. (See Figure \ref{fig:bootstraps}.)

In addition to higher precision at given scan time or shorter scan time at given precision (by up to a factor of four), the optimized sequences display a qualitatively new feature of intrinsic robustness to highly undersampled scans with phase variation such as can arise from magnetic field inhomogeneities. The interplay between these two sources of error has typically resulted in shading artifacts in $T_1$ and $T_2$ maps obtained from low pulse count MRF scans via direct dictionary matching. Alternatives to direct dictionary matching, such as iterative reconstruction \cite{MPM14, ZSA18, ACK17, MWT18} have different characteristics regarding systematic error. In future work, the optimization framework presented here can be applied to these alternative schemes.

To maximize the chance of finding novel pulse sequences, we initialize the optimization algorithms from an ensemble of randomly generated pulse sequences, rather than from an existing human-designed pulse sequence. Over many repetitions from different starting points, the optimization algorithm produces a large number of distinct pulse sequences, which nevertheless consistently display certain qualitative features. Some of these reproduce features which were previously incorporated into MRF pulse sequences designed by human experts while others are qualitatively new features, as discussed in section \ref{sec:discussion} and Figure \ref{fig:stacks}. These new algorithmically-discovered design patterns can inform future MRF pulse sequence design by human experts.

\section{Methods}

\begin{figure}[htbp]
	\begin{center}
	\includegraphics[width=\textwidth]{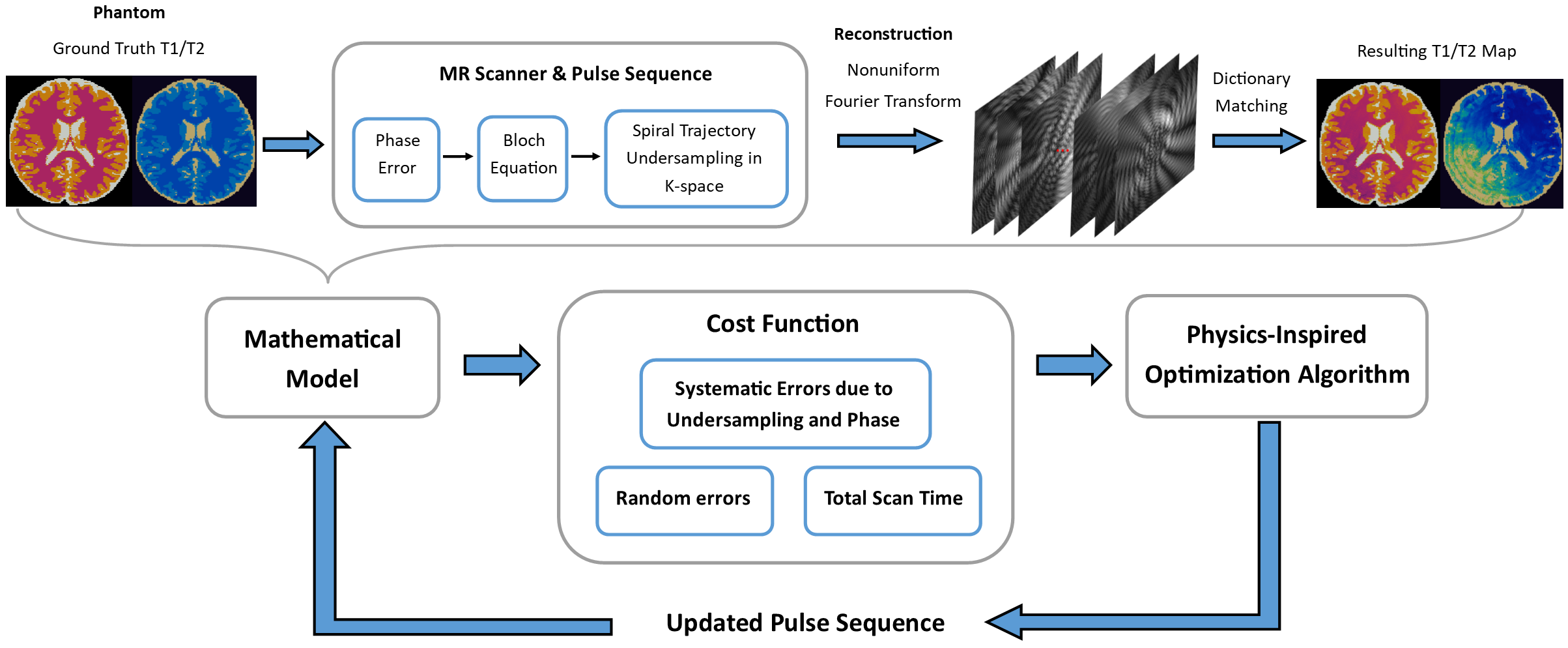} 
	\end{center}
	\caption{\label{fig:flowchart} Here the overall structure of the pulse sequence optimization process is illustrated. A physics-inspired optimization algorithm proposes one or more randomly generated initial pulse sequences, which are then given to a cost function, which returns a quality metric assessing their speed and accuracy. Based on this feedback, the optimization algorithm proposes updated sequences. The cycle of updating and re-evaluation is repeated for a fixed number of iterations. The best sequence found during this process, as judged by the cost function, is produced as final output. Within the cost function, a full simulation of magnetic resonance fingerprinting process is performed. The discrepancy between the simulated ground truth $T_1$ and $T_2$ values in a brain slice and the corresponding values inferred by the standard MRF dictionary-matching procedure are used as a metric of accuracy.}
\end{figure}

We define the sequence optimization problem with three main components: a cost function, a search space of possible pulse sequences, and an optimization algorithm. Figure \ref{fig:flowchart} summarizes the optimization workflow. The theory and implementation of each component are described in the following sections. 

\subsection{MRF pulse sequences}

The specific class of MRF pulse sequences that we consider are Fast Imaging with Steady-state Progression (FISP) sequences \cite{JMS15}, applied in the context of making two-dimensional $T_1$, $T_2$, and proton density maps of the brain\footnote{In this work we ignore the proton density maps inferred from the MRF scans and focus only on the $T_1$ and $T_2$ maps.}. At the start of a FISP pulse sequence, an inversion pulse is applied to initialize the spins of hydrogen nuclei as close as possible to anti-alignment with the background $B_0$ field, which is taken to be in the positive $z$ direction. In such a sequence, step $s$ has total duration $\mathrm{TR}_s$, consisting of the following pieces. First, polar rotation $\alpha_{\tri}$ is applied followed by phase rotation $\theta_{\tri}$. Then, a wait time of $\mathrm{TE}_{\tri} < \mathrm{TR}_{\tri}$ is imposed during which the spins evolve according to exponential decay determined by $T_1$ and $T_2$. Next, the magnetization measurements are performed. Then, further exponential decay occurs during the remaining time $\mathrm{TR}_{\tri} - \mathrm{TE}_{\tri}$. Lastly, a strong ``spoiling'' gradient is applied to mitigate the effects of magnetic field inhomogeneity. (For a sequence diagram see \cite{JMS15}.).

An MRF pulse sequence consists of hundreds or thousands of such steps, with flip angle and TR duration varying from one to the next. In practical settings, MRF scans typically sample Fourier space very sparsely in order to achieve short scan time. Commonly used Fourier undersampling patterns include radial \cite{CKZ16}, spiral \cite{R1}, and 3D Cartesian \cite{KKS20}. Here we consider the ``single-shot'' setting in which one spiral trajectory through $k$-space is sampled after each pulse. The spiral trajectory employed in this study is a variable-density spiral, which needs 48 interleaves to cover $k$-space \cite{R13}. The readout duration is 5.9ms, with a field of view of 300x300 $\textrm{mm}^2$ and matrix size of 256x256 \cite{R1, R13, JMS15, MJC18}. The spiral sampling is varied temporally from one TR to the next, so the aliasing artifacts do not cause constant bias in the signals. A Non-Uniform Fast Fourier Transform (NUFFT) \cite{R14} is then applied to reconstruct images from an MRF scan using measured $k$-space trajectories \cite{DYF98}. Quantification of tissue properties, such as $T_1$ and $T_2$ relaxation times, is performed in each image voxel by comparing the observed magnetization time-series against the entries of a dictionary that contains a list of potential such signals. For each voxel one can assign ($T_1$, $T_2$) based on the dictionary entry that most closely matches the observed signal. Considering system limitations and scan time, we constrain the TR duration to the range of 10 to 100 ms, resulting in a total scan time of on the range of a few seconds for a 2D MRF scan. 3D MRF scans with duration of a few minutes have been reported in \cite{MJD18, CFH20}.

\subsection{Mathematical model}

The dynamics of nuclear spins in a magnetic field are described phenomenologically by the Bloch equation \cite{B46}. Given the magnetic field as a function of time at a given location, an initial condition for all the the spins in a voxel\footnote{To model the effect of spoiling gradients we simulate 400 spins in each voxel. See supporting appendices.}, tissue parameters $T_1$, $T_2$, and acquisition parameters (flip angles, TRs, unbalanced gradients), the Bloch equation predicts the state of the spins at subsequent times. Here, we consider MRF pulse sequences for two-dimensional slices through brain tissue consisting of 256x256 voxels, each 1.2mm by 1.2mm spatial resolution. Correspondingly, our mathematical model consists of a value of $T_1$, $T_2$, and $m_0$ (proton density) assigned to each voxel. The proton density only affects the magnetization of the voxel by acting as a time-independent multiplicative factor. For a given pulse sequence, we solve the Bloch equations (in the hard-pulse approximation) to obtain magnetization vs. time for each ($T_1, T_2$) pair appearing within the voxels of the simulated tissue distribution.

MR scans are complicated processes involving many sources of random and systematic errors. An accurate model of these is a necessary ingredient for an optimization algorithm to produce sequences with good in vivo performance. Here, we model three main sources of error and their interactions: Fourier undersampling artifacts, spatially dependent phase variation, and random error.  Many subtle effects can be present in the interactions. For example, aliasing artifacts are often observed in the $T_1$ and $T_2$ maps derived by applying dictionary matching to Fourier undersampled MRF scans. To minimize these, it is thought to be beneficial to design flip angles and TR times so that the Bloch dynamics spreads signal intensity as uniformly as possible between the sets of TRs associated with each of the $k$-space sampling trajectories. Similarly, shading artifacts are thought to arise through the interplay of two sources of systematic error which individually do not cause shading artifacts, as described below. With a direct first-principles simulation in the cost function, pulse sequences can be designed to suppress such errors without needing to identify and enumerate them.

It is common to observe spatial phase variation in magnetic resonance scans, which could be from spatially varying $B_0$ or $B_1$ field inhomogeneities or temporally varying motion. If not explicitly modeled, such phase variation can contribute a source of systematic error in MRF scans. Although the phase variations observed experimentally vary from scan to scan even on the same machine, they tend to be smoothly varying across the field of view and differ between scans mainly in the direction across which they vary. In our model, we consider static spatially varying phases, which are among the most common sources of systematic error in magnetic resonance scans, and have been reported to cause artifacts and distortion in MR images \cite{LWR16, LKP18}. Furthermore, the interplay between phase variation and aliasing due to Fourier undersampling introduces spatially and temporally varying artifacts that affect both magnitude and phase of the measured signals, which causes a commonly seen ‘shading artifact’ shown in Figure \ref{fig:phase}f from in vivo MRF scans. The shading is not seen when only the undersampling is modeled (Figure \ref{fig:phase}a). Such interactions of artifacts have not been considered in previous work, which is one of the main sources of discrepancy between simulated sequence performance and in vivo scan performance.

In addition to systematic errors, scans exhibit random error, the dominant source of which is Johnson noise in the receive coils. Following \cite{R8, R9}, we model this as independent complex Gaussian zero-mean error at each data point, i.e. white noise.

\subsection{Cost function}

For an MRF scan we wish to minimize $T_1$ error, $T_2$ error, and scan time. We use a weighted combination of the predicted values for these quantities as a cost function to minimize.  Using the formulas from \cite{R8,R9} and an assumed value of variance $\sigma^2$ of a Gaussian noise distribution, one can obtain predicted standard deviation on $T_1$ and $T_2$ for a given tissue due to random errors.  By simulating Fourier undersampling artifacts and phase variation we obtain predictions of discrepancies between the theoretical and measured values of $T_1$ and $T_2$ associated with each voxel. Averaging over voxels of a given tissue type, we can obtain root-mean-square values of systematic error for each of the three tissue types in our model. Correspondingly, an estimate of total error is obtained by taking the sum in quadrature of these root-mean-squared systematic errors with the predicted standard deviations due to random error. We thus obtain six numbers, which are the predicted root-mean-square errors in inferred $T_1$ and $T_2$ for each of the three tissue types in our model (grey matter, white matter, and cerebrospinal fluid).  Ultimately, we wish to minimize these six errors and the total duration of the pulse sequence. Thus we need to combine these seven quantities into a single aggregate measure which our optimization algorithm will attempt to minimize. Optionally, we may add a term incentivizing large average magnetization of the tissues, as this is clearly advantageous for signal to noise ratio.

The cost function $C$ is defined as follows.
\begin{eqnarray}
	C & = & \left( \sigma^{(T_1)} + w_2 \ \sigma^{(T_2)} \right) \sqrt{t} + \frac{ w_{\textrm{mag}} }{\bar{m}_{\min}} \label{cost} \label{eq:cost}\\
	\sigma^{(p)} & = & \sigma^{(p)}_{\mathrm{GM}} + w_{\mathrm{WM}} \ \sigma^{(p)}_{\mathrm{WM}} + w_{\mathrm{CSF}} \ \sigma^{(p)}_{\mathrm{CSF}} \quad \quad p \in \{T_1,T_2\}
\end{eqnarray}

Here, $t$ is the total duration of the sequence, $w_2$ is a tunable ``weight'' quantifying the importance of $T_2$ errors relative to $T_1$ errors, and $w_{\mathrm{WM}}$ and $w_{\mathrm{CSF}}$ are tunable weights quantifying the importance of errors in white matter and cerebrospinal fluid voxels relative to errors in grey matter voxels. $\bar{m}_{\min}$ is magnetization, averaged over TRs and minimized over modeled tissues. $w_{\textrm{mag}}$ is the weight of the incentive on magnetization, which can be set nonzero if a stronger incentive is desired than that provided indirectly through the modeling of random error. The motivation for the form of the $t$-dependence of \eq{cost} is that, by standard sampling statistics, one expects that by using $n$ measurements one can obtain standard deviation scaling as $1/\sqrt{n}$. Consequently, multiplying the total error by $t$ raised to the power $1/2$ or higher should steer the optimizer toward shorter duration sequences. In practice we find that the factor of $\sqrt{t}$ is usually effective. The motivation behind the functional form of the optional magnetization term is that, if magnetization vanishes completely there is no signal and error formally diverges.

Direct evaluation of errors due to Fourier undersampling and phase variation is problematic to incorporate into a cost function due to high computational cost. To speed up evaluation we here introduce a simplifying approximation. Specifically, rather than assigning each voxel to a unique $T_1$, $T_2$ pair, the voxels are all assumed to arise from one of three brain tissue types: grey matter, white matter, and cerebrospinal fluid, each of which has fixed values of $T_1$, $T_2$, and proton density. As a result, the magnetization of the voxels in response to a pulse sequence can be computed by solving the Bloch equations for only three $(T_1, T_2)$ pairs. More importantly, the artifacts due to Fourier undersampling and phase variation can be precomputed based on the spatial distributions of the three tissue types, for each of the 48 $k$-space trajectories. Given these 144 precomputed response functions, a simulation of the reconstructed $T_1$ and $T_2$ maps arising from a 1000 pulse MRF sequence can be completed in 2.0 seconds on a 24 core computer (Azure NC-24 virtual machine). (See Figure \ref{fig:phase} for examples of simulated maps and comparison to in vivo data.)

\subsection{Optimization algorithms}

We formulate the design of MRF pulse sequences as a global optimization program over continuous variables. The cost function is treated as a black box. There is no formula for the gradient of the cost function; strictly speaking, the cost function is not differentiable due to the discrete dictionary matching. Due to the highly non-convex nature of the cost function, we relied on optimization heuristics capable of escaping from local minima. The best performing of these, according to our experimentation, were simulated annealing and substochastic Monte Carlo. The latter is a quantum-inspired optimization algorithm in which quantum fluctuations are modeled in order to escape from local minima. The method used here is a continuous variable generalization of the substochastic Monte Carlo method described in \cite{JJL16}.

Here, we perform optimization with physics-inspired optimization algorithms which have been tuned specifically for the MRF pulse sequence optimization problem. To obtain good pulse sequences, we first reduce the dimension of the search space by only considering pulse sequences which vary smoothly over time. In prior work it has been observed\footnote{An intuitive explanation for this is that, by cycling through different $k$-space trajectories from one TR to the next, one induces undersampling errors that vary in a rapid and discontinuous manner. If these are added to a signal that varies smoothly then this separation in frequency space between signal and noise makes the noise easier to filter out.} that such pulse sequences, when used with dictionary matching, yield $T_1$ and $T_2$ maps with milder Fourier undersampling artifacts than highly discontinuous sequences \cite{R4, R7, STH17, KPK19}. We achieve this by parameterizing the pulse sequences using cubic splines, as has been previously done in other contexts \cite{STH17, HFN15}. The two optimization algorithms found to be most successful (variants of simulated annealing and substochastic Monte Carlo) are both based on biased random walks in the search space. The random perturbations that generate these walks are not only varied in time such that larger, exploratory, perturbations are followed by smaller, fine-tuning, perturbations, but are also non-isotropic, such that different classes of variables are perturbed by different amounts. These non-isotropic updates appear to be crucial to the success of the algorithms, as discussed in the supporting appendices. After a number of runs of the optimization algorithms with different hyperparameters and random seeds, the most promising sequences are selected for in vivo testing.

\subsection{In vivo experiments}

The optimized pulse sequences were tested using in vivo scans to validate our mathematical modeling and directly evaluate the precision, robustness, and image quality of the quantitative tissue maps obtained using optimized MRF pulse sequences. In vivo scans were performed in a Siemens Magnetom Skyra 3T scanner on volunteers following informed consent and approval from the Institutional Review Board.  The scan was acquired with a field of view of 300 x 300 $\textrm{mm}^2$, matrix size of 256 x 256, and slice thickness of 5 mm. As a supplement to subjective judgement of image quality, data from these scans was combined with experimentally measured noise levels in the receive coils to obtain quantitative estimates of random error via bootstrapping statistics, as described in \cite{RBB07}.  The standard deviation in inferred $T_1$ and $T_2$ values from the bootstrap method is calculated within regions of interest in the white matter and is used as a metric of precision.

\section{Results}

\subsection{Optimized pulse sequences}

\begin{figure}[htbp]
	\begin{center}
	\includegraphics[width=\textwidth]{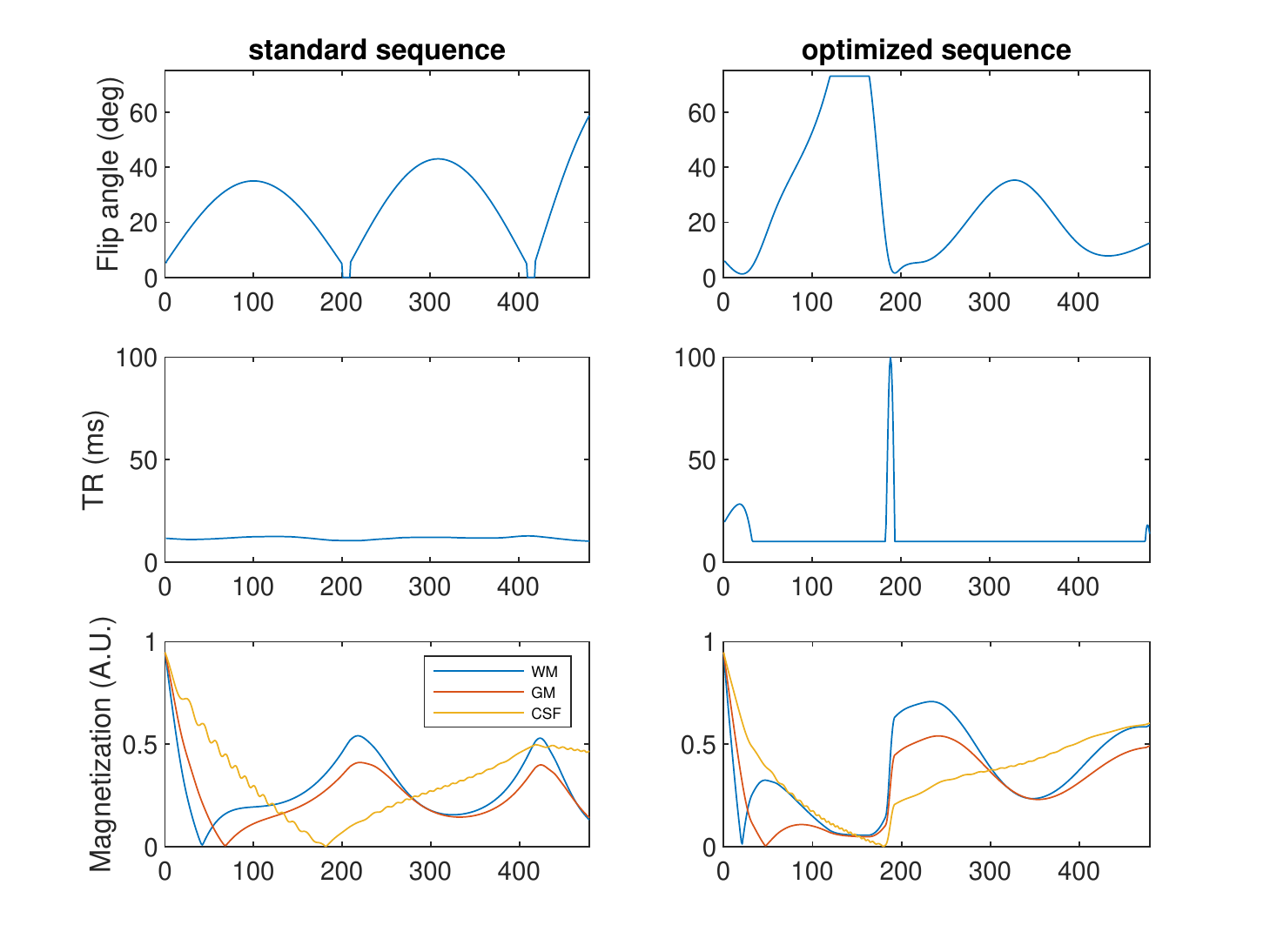} 
	\end{center}
	\caption{\label{fig:stacks} The optimized sequences display qualitatively different features than the standard human-designed sequence. In particular, the optimization algorithm consistently produces pulse sequences in which the TR duration is at its minimum allowed value for most of the TRs, but briefly ``spikes'' to much longer duration. The predicted magnitude of the magnetizations for white matter (WM), grey matter (GM), and cerebrospinal fluid (CSF) are shown for each sequence in units such that the initial inversion pulse achieves magnetization of magnitude 0.95. (The considerations behind modeling the initial magnetization as 0.95 are discussed in \cite{MCC17}.) Optimized sequence $o$ is chosen here as a representative example, about which further information is available in figure \ref{fig:bootstraps} of the main text and tables 1-4 in the supporting appendices. The optimization that produced this sequence used $w_{\mathrm{CSF}} = 0.02452$, $w_{\mathrm{WM}} = 1.000$, $w_2 = 12.02$, and $w_{\mathrm{mag}} = 0$. The pulse sequences are available for non-commercial research purposes from \texttt{https://github.com/madan6711/Automatic-MRF-seq-design}.}
	\end{figure}

Figure \ref{fig:stacks} compares an optimized sequence against a standard MRF scan. The optimized sequences display qualitatively distinct features from prior human-designed sequences, such as ``spiked'' TR durations. Although the details of optimized sequences vary, this spiked TR duration feature is observed consistently for sequences optimized with and without models of phase variation in the cost function. Note that the magnetization curves shown depict the total magnitudes of the magnetization vectors, not just the magnitudes of their projections onto the $xy$-plane, which determine the signal strength of the immediate measurement. However, the latter is strongly sensitive to the flip angle of the given pulse, whereas the former gives a more meaningful metric of the reservoir of available magnetization to be exploited in subsequent pulses. Note also that the optimized sequence shown in Figure \ref{fig:stacks} has total duration only slightly longer than the standard sequence (5.85 seconds vs 5.57 seconds), despite having some TRs that are vastly longer than any TRs in the standard sequence. The optimization algorithm achieves this by setting almost all other TRs to the minimum allowable duration (10 ms), whereas the majority of the TRs in the standard sequence have duration between 11 and 13 ms. 

\subsection{Convergence of optimization algorithms}

\begin{figure}[htbp]
	\begin{center}
	\includegraphics[width=0.65\textwidth]{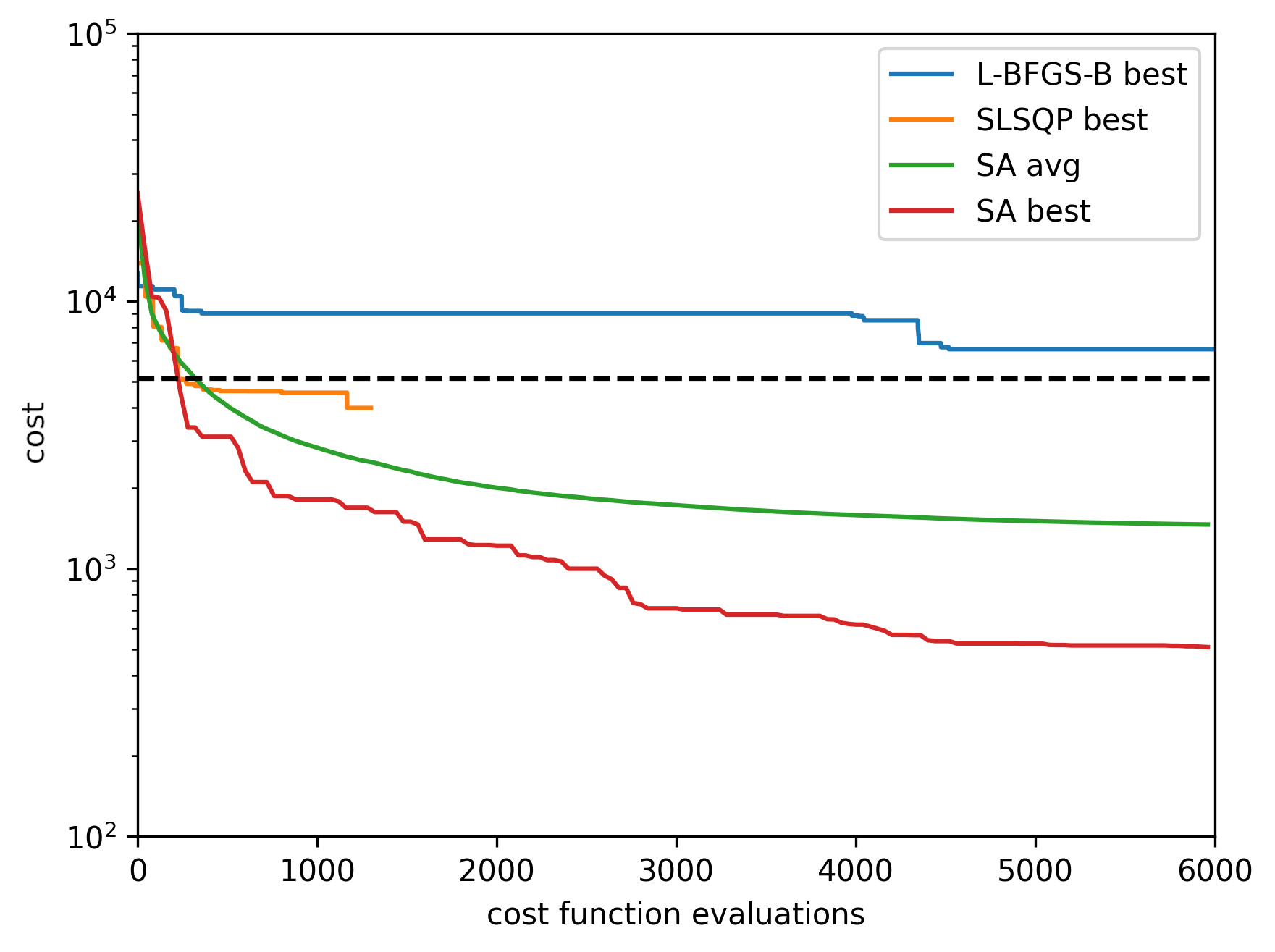}
	\end{center}
	\caption{\label{fig:benchmarking} Comparison of standard optimization routines from SciPy (L-BFGS-B and SLSQP) against our simulated annealing implementation with non-isotropic moves (SA). The decrease in cost is plotted as a function of number of queries made to the cost function. As the evaluation of the cost function is by far the most computationally intensive part of the algorithm, this is therefore a metric of the efficiency of the optimization method. The physics-inspired method's performance varies depending on random seed. Here, the average performance across 200 trials is shown alongside the performance from the best of these trials. (As a meta-algorithm one can run such trials in parallel and select the resulting sequence with lowest cost function value.) For comparison, the cost function value achieved by the standard sequence is shown as a dashed line.}
\end{figure}

Here, we compare three optimization algorithms on the MRF cost function (Eq. \ref{eq:cost}) with a spline-parameterized search space. First, we apply L-BFGS-B, which is a quasi-Newton method which uses gradient information to find local minima of smoothly-varying cost functions \cite{BLN95}. (A variant of BFGS was used previously to optimize MRF pulse sequences in \cite{R7}.) Our cost function is not formally differentiable, due to dictionary matching. Nevertheless, it is smooth on length scales that are not too short, thus yielding meaningful gradient-like information via finite differences. As can be observed from Figure \ref{fig:benchmarking}, the cost function is sufficiently non-convex that, even by starting L-BFGS-B from several random starting points and selecting the best local minimum found yields poor optimization performance. Similarly, we compare to sequential least squares programming (SLSQP) \cite{K88} from multiple random starting points, which also arrives at relatively poor optima upon reaching the algorithm's termination condition. SLSQP is a widely used optmization method closely related to the sequential qudratic programming algorithm employed to optimize MRF pulse sequences in \cite{R4}. The best performance is shown by the physics-inspired Monte Carlo method. Nevertheless, by initiating this with random seeds one never finds precisely the same optima and thus one can conclude that the global optimum is almost certainly not found.

\subsection{Robustness against random error}

\begin{figure}[htbp]
		\begin{center}
			\includegraphics[width=\textwidth]{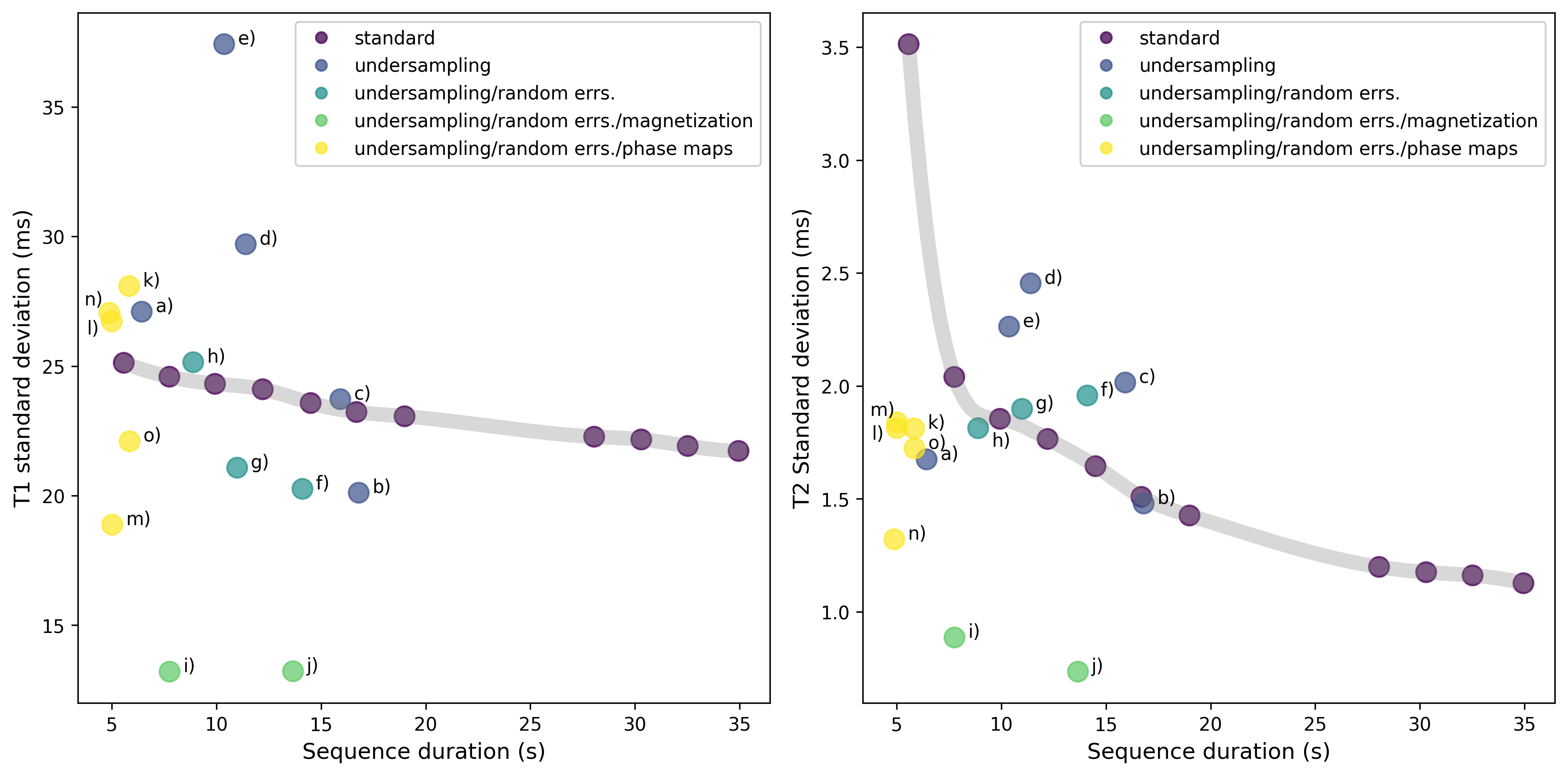}
		\end{center}
	\caption{\label{fig:bootstraps} Precision versus duration tradeoff for optimized and unoptimized sequences. Here the metrics of precision are the standard deviations in inferred $T_1$ and $T_2$ values, which we estimate from in vivo data using bootstrap statistics, as described in \cite{RBB07}. The standard sequence from \cite{R1} is truncated to TR counts from 480 to 3000 in order to obtain scans of different durations, as illustrated by the grey tradeoff curve. The optimized data points are classified according to which terms were included in the cost function. Unsurprisingly, the sequences with best robustness against random error are obtained by heavily incentivizing large signal magnitude (i.e. magnetization) in the cost function. In vivo images corresponding the labelled data points (a-o) are shown in the supporting appendices. Note that these bootstrap statistics are derived from in vivo experimental data for which exact ground truth values of $T_1$ and $T_2$ are inaccessible. Thus they can only assesses scatter and not bias.}
	\end{figure}

Figure \ref{fig:bootstraps} compares the precision of $T_1$ and $T_2$ values between the standard and optimized sequences using various choices of weight parameters $w_2$, $w_{\mathrm{mag}}$, $w_{\mathrm{WM}}$, and $w_{\mathrm{CSF}}$. Unoptimized sequences of different duration are obtained by truncating the sequence from \cite{R1}. These then define a tradeoff curve between precision and duration. As shown in Figure \ref{fig:bootstraps}, the optimizer can yield sequences achieving greater precision at given duration or, equivalently, shorter duration to achieve a given precision, relative to the unoptimized sequences. Specifically, at a given duration, the best optimized sequences achieve over two-fold error reduction in inferred $T_1$ and $T_2$. Alternatively, optimized sequences are found that can achieve over 4.5x speedup relative to unoptimized sequences of comparable precision, as detailed in the supporting information. Note however, that bootstrap statistics measure only random errors and not systematic. Therefore, the optimized sequences which appear best according to bootstrapping may not coincide with the optimized sequences that yield best subjective image quality. 

\subsection{Robustness against systematic error}

\begin{figure}[htbp]
	\begin{center}
	\includegraphics[width=\textwidth]{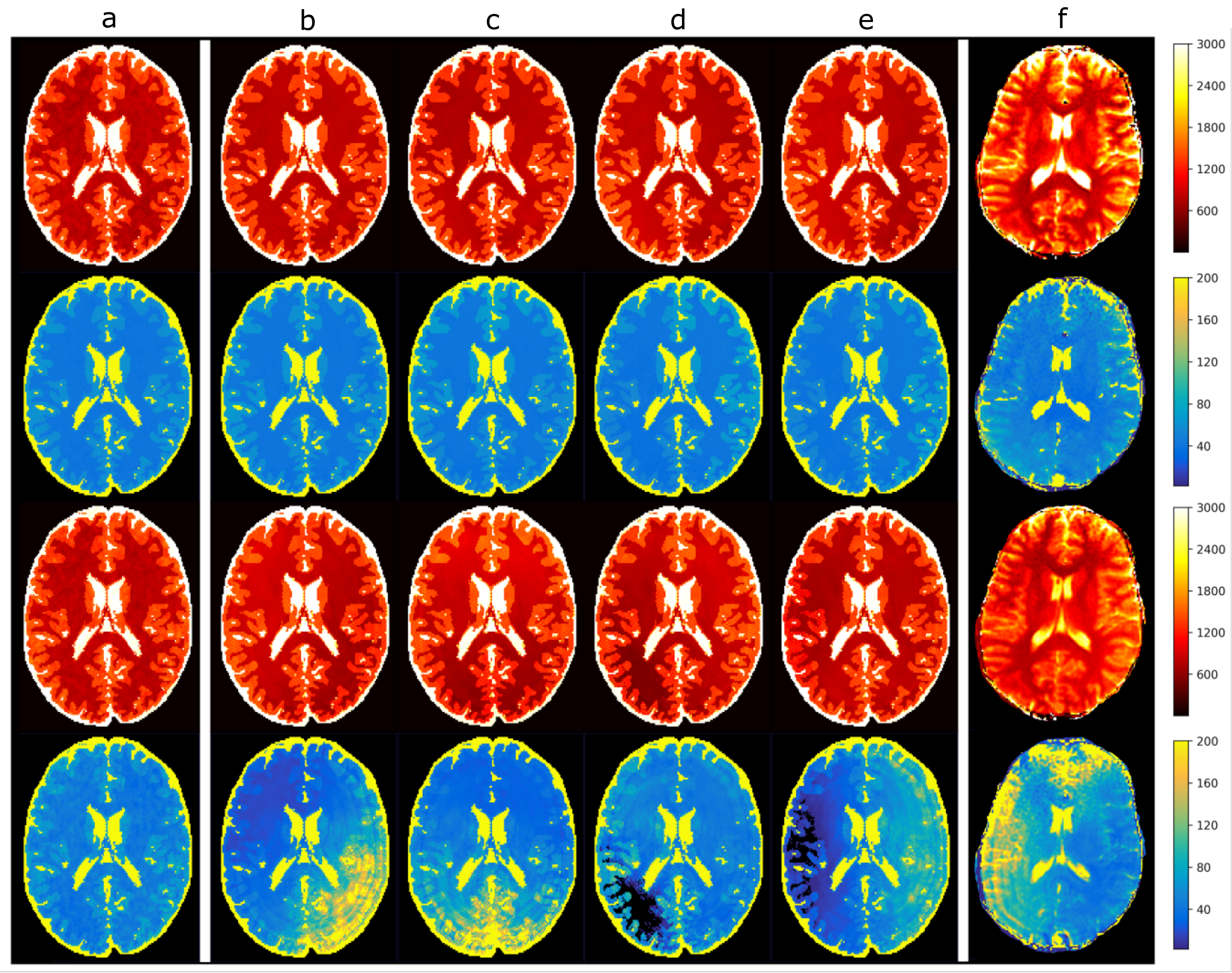}
	\end{center}
	\caption{\label{fig:phase} $T_1$ (red) and $T_2$ (blue) map simulations of an optimized sequence (top two rows) and a standard human-designed sequence \cite{R1} (bottom two rows) incorporating phase variation. The error is modeled as a time-independent phase that varies quadratically along a chosen direction. Experimentally one finds that this direction varies randomly from one scan to the next. In this figure, column \emph{a} includes no phase variation. Columns \emph{b}-\emph{e} correspond to four example orientations for the phase variation. In vivo results for both sequences are shown in column \emph{f}.}
\end{figure}

Robustness against aliasing and systematic errors were estimated in simulations and tested with in vivo scans. Figure \ref{fig:phase} compares the $T_1$ and $T_2$ maps obtained from standard and optimized MRF scans, according to simulation and experiment. In both simulation and experiment, Fourier undersampling artifacts are visible as circular rings and shading artifacts induced by phase variation are visible as non-symmetric image intensity variations. Such errors are quite significant in state of the art short-duration human-designed pulse sequences, as can be seen in the bottom two rows of column \emph{f}.

In addition to in vivo results (column \emph{f}), Figure \ref{fig:phase} shows simulation results under five different assumptions. Column \emph{a} shows simulations incorporating only Fourier undersampling artifacts. The four columns \emph{b}-\emph{e} show simulations that additionally incorporate phase variations in four orientations. (The phase is modeled as varying quadratically across the chosen direction.) Due to the unpredictability of the orientation of phase variation from one scan to the next, maps from simulation do not match in detail the results from individual in vivo scans. However, the simulations which incorporate phase variation accurately predict the relative image quality achieved in vivo by different pulse sequences. Furthermore, the rank-ordering of pulse sequence quality is largely independent of the orientation of the phase homogeneity. Thus high quality (i.e. robust) sequences can be obtained by optimization using a cost function in which a single orientation of variation has been chosen arbitrarily.

It is not a priori obvious that choice of pulse sequence can influence robustness against phase variation. To our knowledge, prior to the algorithmically-discovered pulse sequences shown here, no pulse sequences were known to yield $T_1$ and $T_2$ maps via direct dictionary matching with intrinsic robustness against phase variation.

\section{Discussion}
\label{sec:discussion}

Here, we present an automatic pulse sequence design framework for MR Fingerprinting scans using physics-inspired optimization.  The cost function is built upon explicit first-principles modeling of MRF scans, incorporating random error as well as errors induced by the interplay of Fourier undersampling, phase variation, and image reconstruction algorithms. This realistic modeling, combined with tailored optimization algorithms, yields novel MRF pulse sequences that strongly outperform standard MRF pulse sequences according to in vivo experiments.

Although the optimization algorithms producing these MRF pulse sequences are essentially inscrutable, by examining the qualitative features of the optimized sequences one may obtain insights that can be fed back to human-led MRF pulse sequence design efforts. The first of these is that careful design of MRF pulse sequences can greatly reduce susceptibility to shading artifacts. This is an important finding in this study, because the shading artifacts represent a unique interaction of static phase variation, spatially and temporally varying undersampling errors, and signal intensity variation from temporally varying flip angles and timings. Second, the flip angle versus pulse index in high-performing sequences is consistently observed to consist of a modest number (3-6) of ``humps.'' This is a feature present in prior MRF pulse sequences designed by human experts. That optimization algorithms reproduce it when proceeding from randomly generated starting points is strong confirmation of the intuitions behind this design. Third, plots of TR duration, (i.e. time between pulses), vs. pulse index arising from optimized sequences consistently show spike patterns. That is, all but a small number of the TRs are of minimum allowed duration, with a small number of TRs with vastly longer duration.

We propose the following interpretation for why the optimizer finds spiked TR patterns to be beneficial. Rapidly gathering data that efficiently distinguishes tissues is aided by short TR times and large flip angles. However, repeated application of pulses with these features gradually depletes magnetization levels and hence signal strength. One might interpret the TR duration spikes as ``rest periods'' during which data collection is temporarily sacrificed in favor of allowing $T_1$ relaxation to proceed unimpeded and thereby ``recharge'' magnetization levels. This interpretation is bolstered by observation that, across large ensembles of optimized sequences, the TR duration spikes are generally coordinated with low flip angles, as illustrated in figure 10 of the supporting appendices. This is a qualitatively new feature absent in prior human-designed sequences and may be a significant contributor to the improved duration versus accuracy tradeoff. 

Our optimized sequences achieve increased scan speed at a given precision target relative to standard sequences and can simultaneously yield intrinsic robustness agaist phase variation. Although our in vivo scans were performed on a scanner with 3 Tesla field strength, optimization of sequences to achieve robustness against systematic variations may in future work serve as an enabling technology for  low-cost low-field-strength portable scanners, which are likely to have large field inhomogeneities and low signal to noise ratio.

In this study, we modeled three types of errors that are commonly seen in the in vivo MRF scan. In addition to Gaussian random noise, we explicitly modeled 2D spatially and temporally dependent artifacts due to undersampling and phase variation. The need to incorporate phase variation is a unique finding in this study. In a fully sampled MRF scan, this phase variation is time independent and would have no effect on the maps derived from dictionary matching. However, in an actual in vivo MRF scan which typically employs high acceleration rate and time-varying sampling trajectories, the phase variation combine with Fourier aliasing to generate spatially and temporally varying artifacts that cause ‘shading’ artifacts in the maps. Figure 5 compares simulation with and without incorporating phase variation. Only the simulation incorporating phase variation reproduces ‘shading’ from the in vivo scan. Due to the interplay of multiple error sources, the optimal MRF sequence design should be a comprehensive consideration of flip angle series, timing, sampling trajectories and reconstruction.

Our cost function is based on a digital brain phantom from the Montreal Neurological Institute Brain Imaging Center \cite{AEC06, AGP06}. For practical implementation we simplified the model by mapping all voxels to three tissue types: grey matter, white matter, and cerebrospinal fluid. Although optimization using this limited model yielded sequences with good in vivo performance, one could in future work consider more detailed models involving additional tissue types and geometries, including pathological cases. Furthermore, with enough computational resources, optimization could be performed using a cost function that incorporates a complete non-uniform Fourier Transform rather than precomputed response functions. This would enable $k$-space trajectories to be co-optimized with the other acquisition parameters such as flip angles and TR durations. 

The general framework developed here for automated design of pulse sequences via physics-inspired optimization algorithms could be applied in many other contexts. The framework would be straightforward to adapt to optimize pulse sequences for three-dimensional scans, or for operation with alternative image reconstruction schemes such as iterative reconstruction. Furthermore, the optimization algorithms, parameterization of the search space, and cost function described here could be adapted to the development of MRF pulse sequences specialized for body parts other than the brain, as well as for MRF scans measuring quantities beyond $T_1$ and $T_2$ such as diffusion rates. More ambitiously, because our cost function explicitly models the systematic errors arising from specific tissue distributions, it could be used to generate MRF pulse sequences tailored to specific disorders or even specific patients. In the opposite direction, repeatability of pulse sequence results across different scans and different patients could be addressed by large scale in vivo studies. Such repeatability is an especially important feature for monitoring disease progression or age related tissue changes \cite{BYR15, WBG18, FBC20} and helps to fulfil the promise of quantitative MR scanning for objective clinical diagnostic criteria \cite{FBC20, KKL19}.

\noindent \textbf{Acknowledgments:} This work was partially supported by Siemens Healthineers and by the National Institutes of Health through grants NIH R21EB026764-01 and NIH R01NS109439-01.

\begin{appendices}

\section{Spin Dynamics}

The dynamics of nuclear spins in a magnetic field are described phenomenologically by the Bloch equation \cite{B46}. Given the magnetic field as a function of time at a given location, an initial condition for the spin, and parameters $T_1$, $T_2$, the Bloch equation predicts the state of the spin at subsequent times. Here, we consider MRF pulse sequences for two-dimensional slices through brain tissue consisting of $256 \times 256$ voxels, each 1.2mm by 1.2mm. (The thickness is not explicitly modeled but in vivo is approximately 5mm.) Correspondingly, our mathematical model consists of a value of $T_1$, $T_2$, and $m_0$ (proton density) assigned to each voxel. The proton density only affects the magnetization of the voxel by acting as a time-independent multiplicative factor. For a given pulse sequence we solve the Bloch equations (in the hard-pulse approximation) to obtain magnetization vs. time for each $(T_1, T_2)$ pair appearing within the voxels of the simulated tissue distribution.

Because the static $B_0$ field is not perfectly spatially homogeneous, different spins throughout the brain precess at slightly different rates. This induces unwanted artifacts in the resulting magnetic resonance images. The purpose of the spoiling gradient is to reduce sensitivity to $B_0$ inhomogeneity by effectively averaging away the $x$ and $y$ components of the magnetization at the end of each TR. For this to be effective, the spoiling gradient needs to be sufficiently strong that the difference in precession angle between different spins within the same voxel is at least $2 \pi$.  In our computer model we assign $F = 400$ spins to each voxel, which get rotated by angles uniformly spaced between $-\pi$ and $\pi$ during the spoiling gradient. At the measurement stage, the magnetization associated to a given voxel is obtained by averaging over these 400 spins, conventionally referred to as isochromats.

In the hard pulse approximation, the RF pulses that rotate the spins on the Bloch sphere are considered instantaneous. Consequently, the exponential decay dictated by $T_1$ and $T_2$ is not intermixed with these rotations. Let $\tri$ denote the index of a given TR. The first step in a given TR is to apply an RF pulse implementing a rotation on the Bloch sphere according to polar angle $\alpha_s$ followed by azimuthal angle $\theta_s$. That is, the magnetization vector $\vec{m}_{\tri,j}$ of the $j\th$ isochromat at the $\tri\th$ timestep undergoes the transformation
\begin{equation}
    \vec{m}_{\tri,j}(x,y) \gets R_{\tri} \ \vec{m}_{\tri,j}(x,y) 
\end{equation}
where $R_\tri$ is the rotation matrix
\begin{equation}
    R_{\tri} = \left[ \begin{array}{ccc} \cos^2 \theta_{\tri} + \cos \alpha_{\tri} \sin^2 \theta_{\tri} & (\cos \alpha_{\tri} - 1) \cos \theta_{\tri} \sin \theta_{\tri} & \sin \alpha_{\tri} \cos \theta_{\tri} \\
        (\cos \alpha_{\tri} - 1) \cos \theta_{\tri} \sin \theta_{\tri} & \cos \alpha_{\tri} \cos^2 \theta_{\tri} + \sin^2 \theta_{\tri} & \cos \theta_{\tri} \sin \alpha_{\tri} \\
        - \sin \alpha_{\tri} \sin \theta_{\tri} & - \cos \theta_{\tri} \sin \alpha_{\tri} & \cos \alpha_{\tri} \end{array} \right].
\end{equation}
The next step in the $\tri\th$ TR is to wait for time $\mathrm{TE_s}$. Left undisturbed for duration $\mathrm{TE}_s$ the magnetization will relax toward equilibrium according to
\begin{equation}
    \label{eq:relaxation}
    \vec{m}_{\tri,j}(x,y) \gets D(\mathrm{TE}_{\tri}) \ \vec{m}_{\tri,j}(x,y) + \vec{v}(\mathrm{TE}_{\tri}).
\end{equation}
Where
\begin{equation}
    D(t) = \left[ \begin{array}{ccc} e^{-t/T_2} & 0 & 0 \\
        0 & e^{-t/T_2} & 0 \\
        0 & 0 & e^{-t/T1} \end{array} \right]
\end{equation}
and
\begin{equation}
    \vec{v}(t) = \left[ \begin{array}{c} 0 \\ 0 \\ 1-e^{-t/T_1} \end{array} \right].
\end{equation}
Conventional magnetic resonance imaging hardware cannot measure the $z$-component of magnetization. Furthermore, the measurement cannot distinguish isochromats within a voxel but instead is sensitive only to their average magnetization. In keeping with widely used conventions in the magnetic resonance literature we express this average magnetization in the $xy$ plane in a given voxel as a complex number whose real part is the $x$-component of the magnetization and whose imaginary part is the $y$-component, as follows.
\begin{equation}
    m_{\tri}(x,y) \gets \frac{1}{F} \sum_{j = 0}^{F-1} \left( [\vec{m}_{\tri,j}(x,y)]_x + i \ [\vec{m}_{\tri,j}(x,y)]_y \right).
\end{equation}
Next, another idle waiting period is imposed for the remaining time $\mathrm{TR}_\tri - \mathrm{TE}_\tri$. Hence, relaxation dynamics again occurs in accordance with \eq{eq:relaxation}.
\begin{equation}
    \vec{m}_{\tri,j}(x,y) \gets D(\mathrm{TR}_{\tri} - \mathrm{TE}_{\tri}) \ \vec{m}_{\tri,j}(x,y) + \vec{v}(\mathrm{TR}_{\tri} - \mathrm{TE}_{\tri}).
\end{equation}
Lastly, a spoiling gradient is applied, which rotates the different isochromats within the voxel by different angles, determined by their position along the gradient. That is,
\begin{equation}
    \vec{m}_{\tri+1,j}(x,y) \gets S_j \ \vec{m}_{\tri,j}(x,y),
\end{equation}
where
\begin{equation}
    S_j = \left[ \begin{array}{ccc} \cos \phi_j & -\sin \phi_j & 0 \\
        \sin \phi_j & \cos \phi_j & 0 \\
        0 & 0 & 1 \end{array} \right].
\end{equation}
This mathematical model of a FISP pulse is summarized in figure \ref{fig:bloch_summary}.

For a sequence with an initial inversion pulse we take the initial state of $\vec{m}_j$ to be $(0,0,-0.95)$ for all $j$. 

\begin{figure}
\[
\begin{array}{rcll}
\vec{m}_{\tri,j}(x,y) & \gets & R_{\tri} \ \vec{m}_{\tri,j}(x,y) & \textrm{apply rotation specified by $(\alpha_{\tri}, \theta_{\tri})$}\\
\vec{m}_{\tri,j}(x,y) & \gets & D(\mathrm{TE}_{\tri}) \ \vec{m}_{\tri,j}(x,y) + \vec{v}(\mathrm{TE}_{\tri}) & \textrm{wait for time $\mathrm{TE}_{\tri}$}\\
m_{\tri}(x,y) & \gets & \frac{1}{F} \sum_{j = 0}^{F-1} \left( [\vec{m}_{\tri,j}(x,y)]_x + i \ [\vec{m}_{\tri,j}(x,y)]_y \right) & \textrm{measurement averages isochromats} \\
\vec{m}_{\tri,j}(x,y) & \gets & D(\mathrm{TR}_{\tri} - \mathrm{TE}_{\tri}) \ \vec{m}_{\tri,j}(x,y) + \vec{v}(\mathrm{TR}_{\tri} - \mathrm{TE}_{\tri}) & \textrm{wait for remainder of $\mathrm{TR}_{\tri}$ duration} \\
\vec{m}_{\tri+1,j}(x,y) & \gets & S_j \ \vec{m}_{\tri,j}(x,y) & \textrm{apply spoiling gradient}
\end{array}
\]
\caption{\label{fig:bloch_summary}Summary of spin dynamics within a voxel of given $T_1$ and $T_2$ during the $\tri\th$ TR of a FISP pulse sequence.}
\end{figure}

\section{Digital Phantom}
\label{sec:phantom}

For an MRF scan we wish to minimize $T_1$ error, $T_2$ error, and scan time. We use a weighted combination of the predicted values for these quantities as a cost function to minimize. Magnetic resonance scans are complicated processes involving many sources of random and systematic error. Modeling these to obtain a cost function that accurately predicts real-world in vivo performance of pulse sequences is highly nontrivial. Here, we use a model that incorporates random error due to thermal noise, and systematic errors due to Fourier undersampling and phase inhomogeneity. Such phase inhomogoneity is commonly observed and could be caused by $B_0$ or $B_1$ inhomogeneity or motion. (A preliminary report on our cost function appears in \cite{R12}).

At least ideally, the RF pulses in a magnetic resonance scan are uniform throughout the $xy$ plane and induce the same rotation on the Bloch sphere to every spin throughout the targeted slice of tissue. Sensitivity to spatial variation is obtained during the measurement step through the use of magnetic field gradients, which allow the extraction of Fourier components of the magnetization in the $xy$-plane. Because spins at different locations have different values of $T_1$ and $T_2$ depending on tissue type, the magnetization vs time will vary spatially, and this variation can be used to map the distribution of different tissues. To obtain complete information about the magnetization after each pulse, one would need to measure a number of Fourier components equal to the number of voxels in the desired image. However, to obtain shorter scan durations it is typical in practice to measure only a much smaller number of Fourier components after each pulse.

In this study we use a high level of Fourier undersampling to achieve short scan times. Specifically, we use a variable density spiral readout in Fourier space \cite{R13} in the ``one-shot'' setting with undersampling factor $R=48$. That is, after each RF pulse, we measure Fourier components along a single spiral, repeatedly cycling through a sequence of 48 spirals, which collectively provide full coverage of Fourier space. Superimposing all 48 spirals does not yield a set of points in Fourier space arranged according to a uniform grid. Rather, the density of samples in Fourier space varies slightly. To infer spatial images, this non-uniform sample density is compensated for using a Non-Uniform Fast Fourier Transform (NUFFT) \cite{R14}.

The mapping from actual $xy$-magnetization at a given time to the measured Fourier components is given by a Fourier transform, which is linear. The mapping from the measured Fourier components to the inferred $xy$-magnetization is given by a non-uniform Fourier Transform, which is also linear. Therefore, the inferred magnetization in position space is expressible as a linear combination of contributions from the actual magnetization in position space. That is,
\begin{equation}
\label{full_psf}
I_{\tri}(x,y) = \sum_{x',y'} U_{\tri}(x,y,x',y') m_{\tri}(x',y').
\end{equation}
Here, $m_{\tri}(x',y')$ is the actual magnetization at location $(x',y')$ at the time of $\tri\th$ measurement, $I_{\tri}(x,y)$ is the magnetization at location $(x,y)$ inferred based on the results of the $\tri\th$ measurement, and $U_{\tri}(x,y,x',y')$ is the point spread function defined by the non-uniform Fourier transform applied to the set Fourier components measured in the $\tri\th$ step. (In our case, the set of Fourier components measured in the $\tri\th$ step are those lying within the $j\th$ spiral trajectory, where $j$ is given by $\tri$ reduced modulo 48. Thus $U_{\tri}(x,y,x',y') = U_{\tri+48}(x,y,x',y')$.) Here we are taking $I_{\tri}(x,y)$ and $m_{\tri}(x',y')$ to be complex numbers as noted earlier.

In magnetic resonance fingerprinting, one discretizes the range of $T_1$ and $T_2$ that might be found in human tissues into a finite set of values. Given a pulse sequence, one then computes, for each $(T_1,T_2)$ pair in this set, the corresponding magnetization vs. measurement index $\tri$. This list of potential magnetization vs. $\tri$ curves is called a dictionary. After performing a magnetic resonance scan and applying non-uniform Fourier transforms, one obtains estimates of magnetization vs. measurement index for each voxel. Although the individual magnetization estimates for each measurement index have large error due to Fourier undersampling, the time-series across all measurement indices nevertheless contains useful information. The time series for a given voxel can be compared against the entries in the dictionary, and for each voxel one can assign $(T_1,T_2)$ based on the dictionary entry that makes the closest match to the observed signal according to a suitably chosen metric of closeness. Here, following \cite{R1}, we take the perspective that, for a sequence with $n$ measurements, we can normalize $(I_1(x,y), I_2(x,y),\ldots,I_n(x,y))$ to obtain a unit vector in $\mathbb{C}^n$. The dictionary entries are also normalized to become unit vectors in $\mathbb{C}^n$. For a given voxel one infers $(T_1,T_2)$ to be the values of the dictionary entry whose inner product with $(I_1(x,y), I_2(x,y),\ldots,I_n(x,y))$ has the largest magnitude.

In principle, given a pulse sequence, choice of Fourier-space trajectories, and a model tissue distribution assigning $(T_1,T_1,m_0)$ values to each voxel, one can solve the Bloch equation to obtain $m_{\tri}(x',y')$ and solve \eq{full_psf} to obtain $I_{\tri}(x,y)$, thereby simulating the effect of Fourier undersampling errors. This data can then be matched against a dictionary of predicted signals to obtain inferred $(T_1,T_2)$ values for each voxel. The inferred $(T_1,T_2)$ values can then be compared against the model tissue distribution to evaluate error in inferred $T_1$ and $T_2$ induced by Fourier undersampling. Some pulse sequences will be more robust against Fourier undersampling error than others, and this metric of error can thus be used to construct a cost function to optimize.

Unfortunately, this is not very practical, as most optimization methods need to make a large number of queries to the cost function and evaluation of $I_{\tri}(x,t)$ via \eq{full_psf} is somewhat computationally intensive. Large computational savings can be made by taking a somewhat simplified model, as described in \cite{R12}. Instead of assigning each voxel in the model brain to a unique $(T_1,T_2,m_0)$ value, we consider an idealized brain in which each voxel in our $256 \times 256$ array is one of four types: white matter ($T_1 = 800\mathrm{ms}, T_2 = 40\mathrm{ms}, m_0 = 0.77$), grey matter ($T_1 = 1400\mathrm{ms}, T_2 = 60\mathrm{ms}, m_0 = 0.86$), cerebrospinal fluid ($T_1 = 3000\mathrm{ms}, T_2 = 2000\mathrm{ms}, m_0 = 1.0$), or air ($m_0 = 0$). Consequently, to compute $I_{\tri}(x,y)$, one need only to solve the Bloch equations for grey matter, white matter, and cerebrospinal fluid, and then for each measurement index $\tri$, take the corresponding complex linear combination of the three pre-summed point spread functions corresponding to the spatial distributions of these three tissues. That is,
\begin{equation}
\label{3tissue}
I_{\tri}(x,y) = U_{\tri}^{(\mathrm{GM})}(x,y) m_{\tri}^{(\mathrm{GM})} + U_{\tri}^{(\mathrm{WM})}(x,y) m_{\tri}^{(\mathrm{WM})} + U_{\tri}^{(\mathrm{CSF})}(x,y) m_{\tri}^{(\mathrm{CSF})}
\end{equation}
where:
\begin{equation}
\label{ui}
U_{\tri}^{(\mathrm{GM})}(x,y) = m_0^{(\mathrm{GM})} \sum_{(x',y') \in \mathrm{GM}} U_{\tri}(x,y,x',y')
\end{equation}
and similarly for WM and CSF.

Although spoiling gradients, as used in FISP sequences, mitigate the effects of $B_0$ inhomogeneity, they do not eliminate spatial inhomogeneities entirely. It has been previously reported in highly undersampled MRF scans that systematic errors in the phase of $m_{\tri}(x,y)$ result in shading artifacts in the inferred $T_1$ and $T_2$ images, even in FISP sequences \cite{KPK19}. Examples of such shading artifacts are shown in figure 1 of the main text. Currently, this is usually dealt with in postprocessing, by employing methods such as iterative reconstruction \cite{MPM14, ZSA18, ACK17, MWT18}. Here, we take a novel approach of optimizing pulse sequences to be intrinsically robust against these errors, thereby producing good quality images directly from inner-product maximizing dictionary matching, without the need for ad hoc corrections.

To optimize for robustness against phase errors we can simply incorporate a representative example of typically observed phase errors into the point spread functions $U_{\tri}^{(\mathrm{GM})}(x,y)$, $U_{\tri}^{(\mathrm{WM})}(x,y)$, and $U_{\tri}^{(\mathrm{CSF})}(x,y)$. The phase errors observed experimentally vary from scan to scan even on the same machine. However, the phase errors tend to be smoothly varying across the field of view and differ between scans mainly in the direction across which they vary.  Empirically, as discussed in the results section, we have found that optimizing against a representative example of a phase error tends to yield sequences that are also robust against other phase errors with different orientations. Furthermore, the sequences with smaller shading artifacts in simulation are observed to have smaller shading artifacts in vivo. Examples of simulated phase errors are given in figure 2 of the main text.

After the predicted magnetization time-series are predicted for each voxel, these are checked against a ``dictionary'' of simulated time-series for a list of possible $T_1,T_2$ pairs. The value of $T_1$ and $T_2$ from the dictionary entry that most closely matches the measured time signal from a given voxel are inferred as the most likely estimates of the ground truth $T_1$ and $T_2$ values for that voxel.  In this manner a maps of $T_1$ and $T_2$ are extracted, in keeping with standard practice in Magnetic Resonance Fingerprinting \cite{R1}. At each query, the cost function is given a new pulse sequence, and thus must generate a new MRF dictionary. In our modeling we use a dictionary of 14,996 $(T_1,T_2)$ pairs, discretizing the range
\begin{equation}
\begin{array}{rcccl}
2\mathrm{ms} & \leq & T_1 & \leq & 3000\mathrm{ms} \\
2\mathrm{ms} & \leq & T_2 & \leq & 2000\mathrm{ms} \\
& T_2 & \leq & T_1. &
\end{array}
\end{equation}
Thus, at each query to the cost function, the Bloch equations must be solved for each of these 14,996 $(T_1,T_2)$ values. Then, after the Fourier undersampling errors have been simulated according to \eq{3tissue}, the inner products between the signals calculated for each of the $256 \times 256$ voxels (with the exception of the air voxels) must be calculated with the signals calculated for each of the 14,996 library entries in order to perform the dictionary decoding. These two processes: dictionary generation, and dictionary decoding, are the dominant computational costs in the evaluation of the cost function, with the evaluation of \eq{3tissue} and the evaluation of of the predicted random errors being essentially negligible. Using an optimized multithreaded implementation running on a 24-vcpu virtual machine (Azure NC24) we find that the cost function, for a sequence with 1000 TRs, can be evaluated in 2.0 seconds.

It is unlikely that any mathematical model of a complicated system such as this one will ever be complete. In particular, our digital phantom does not include explicit modeling of slice profile corrections. It also does not directly model of spatial inhomogeneity of $B_0$ or $B_1$ magnetic fields or time-dependent effects due, for example, to eddy currents. Rather, these are incorporated via a simple phenomenological model in which we impose a phase that varies quadratically along an arbitrary direction in the xy-plane. The merits of this model are that it can be computed rapidly enough to incorporate into a cost function and it empirically does a good job of qualitatively reproducing the systematic errors observed in a series of in vivo scans that were carried out using a wide variety of pulse sequences on several volunteers. In future work, one could consider incorporating first-principles modeling of additional physical effects into the digital phantom.

\section{Model of Random Errors}

In addition to Fourier undersampling and phase errors, magnetic resonance scans are also affected by random error due to thermal fluctuations. These are typically modeled as independent identically distributed gaussian errors of mean zero and variance $\sigma^2$ added to each measured Fourier coefficient. Analytical formulas for the resulting errors in inferred $T_1$ and $T_2$ via dictionary matching are derived in \cite{R8, R9}. In addition to $\sigma^2$, these errors depend on the rate at which the dictionary entries vary with respect to $T_1$ and $T_2$. These formulas show that better robustness is achieved by pulse sequences such that the dictionary entries (thought of as vectors in $\mathbb{C}^n$) vary rapidly as $T_1$ and $T_2$ are changed. This is in agreement with general intuition. In fact, prior work has used small inner product between adjacent dictionary entries as a criterion for optimizing pulse sequences \cite{R10}.

Using the formulas from \cite{R8, R9} and a value of $\sigma^2$, one can obtain predicted standard deviation on $T_1$ and $T_2$ for a given tissue. The value of $\sigma^2$ can either be inferred from experimental data or treated as effectively a tunable ``weight'' factor to adjust the importance of random error relative to systematic error in the optimization. From our simulation of Fourier undersampling artifacts we obtain predicted discrepancies between the theoretical and measured values of $T_1$ and $T_2$ associated with each voxel. Averaging over voxels of a given tissue type, we can obtain root-mean-square values of systematic error due to Fourier undersampling for each of the three tissue types in our model. These can be interpreted as standard deviations if one were to select a voxel uniformly at random among all voxels of the given tissue type. Correspondingly, we add these root-mean-squared undersampling errors to the predicted standard deviations to obtain a total predicted error. We thus obtain six numbers:
\begin{equation}
    \label{eq:err1}
    \sigma^{(p)}_{t} = \sqrt{ \left(\nu^{(p)}_{t}\right)^2 + \left(\eta^{(p)}_{t}\right)^2} \quad p \in \{T_1,T_2\} \quad t \in \{ \mathrm{GM}, \mathrm{WM}, \mathrm{CSF} \},
\end{equation}
where $\nu^{(p)}_{t}$ is the standard deviation in parameter $p$ and tissue $t$ predicted due to thermal noise, and $\eta^{(p)}_{t}$ is the root-mean-square error in parameter $p$ and tissue $t$ predicted due to Fourier undersampling.

The sensitivity to thermal noise is affected by the choice of pulse sequence through several mechanisms. First, a sequence that yields larger magnetization in the $xy$-plane will yield stronger signals and hence better signal to noise ratio. More subtly, some sequences are better than others in terms of how rapidly the dictionary entry (which for a scan with $n$ measurements can be thought of as a vector in $\mathbb{C}^n$) varies as a function of $T_1$ and $T_1$. If the dictionary entry varies rapidly as a function of these parameters then the inferred values of these parameters will be less affected by random error in the measured signal. A mathematical analysis of this effect is given in \cite{R8,R9}.

Although we have motivated the above error model heuristically, it is worth highlighting that it can be derived from a minimal set of assumptions, which do not include any assumption about errors being Gaussian. Specifically, let $\delta_{j,t,p}$ be the (signed) discrepancy between the true value of $p \in \{T_1,T_2\}$ for voxel $j$ of tissue $t \in \{\mathrm{GM}, \mathrm{WM}, \mathrm{CSF}\}$ and the value inferred by dictionary matching. Then, by definition, the mean and variance of $\delta_{j,t,p}$ are
\begin{eqnarray}
    \mu_{j,t,p} & = & \langle \delta_{j,t,p} \rangle \label{eq:mu} \\
    \nu^2_{j,t,p} & = & \langle \delta_{j,t,p}^2 \rangle - \langle \delta_{j,t,p} \rangle^2, \label{eq:sigma2}
\end{eqnarray}
respectively. Let $N_p$ be the number of voxels of tissue type $p$. If we select a voxel uniformly at random among these voxels then the root-mean-squared error is\footnote{By linearity of expectation one could equivalently write $\sigma_t^{(p)} = \sqrt{\langle \frac{1}{N_p} \sum_{j=1}^{N_p} \delta_j^2 \rangle}$.}
\begin{equation}
    \sigma_t^{(p)} = \sqrt{\frac{1}{N_p} \sum_{j=1}^{N_p} \langle \delta_j^2 \rangle}
\end{equation}
By \eq{eq:mu} and \eq{eq:sigma2}
\begin{equation}
    \sigma_t^{(p)} = \sqrt{ \frac{1}{N_p} \sum_{j=1}^{N_p} \left( \nu^2_{j,t,p} + \mu^2_{j,t,p} \right) } \label{eq:penultimate_error}
\end{equation}
In our error model, $\mu_{j,t,p}$ is calculated separately for each voxel by explicitly modeling the Fourier undersampling and phase errors, applying dictionary matching to infer the value of parameter $p$ for voxel $j$, and then subtracting from that the original ground truth value of parameter $p$ for voxel $j$ in the original model. In contrast, we estimate $\nu^2_{j,t,p}$ for each tissue type and parameter using the perturbative arguments of \cite{R8, R9}. Thus our estimated values of $\nu^2_{j,t,p}$ are in fact independent of $j$. Consequently, \eq{eq:penultimate_error} simplifies to
\begin{equation}
    \label{eq:almost}
    \sigma_t^{(p)} = \sqrt{ \nu^2_{t,p} + \frac{1}{N_p} \sum_{j=1}^{N_p} \mu^2_{j,t,p} }.
\end{equation}
Introducing the notation $\eta_t^{(p)} = \sqrt{\frac{1}{N_p} \sum_{j=1}^{N_p} \mu^2_{j,t,p}}$ for the root-mean-square systematic error and substituting into \eq{eq:almost} yields \eq{eq:err1}. Note also that there is no assumption that error is of mean zero. Bias is incorporated into the error metric via the $\left( \eta_t^{(p)} \right)^2$ term.

\section{Magnitude Incentive}
\label{sec:magnitude}

As shown in equation (1) of the main text, our cost function takes the form
\begin{equation}
    \label{eq:multi}
    C = C_{\mathrm{main}} + w_{\mathrm{mag}} C_{\mathrm{mag}},
\end{equation}
where
\begin{equation}
    C_{\mathrm{main}} = \left( \sigma^{(T_1)} + w_2 \sigma^{(T_2)} \right) \sqrt{t}
\end{equation}
and
\begin{equation}
    \label{eq:cmag}
    C_{\mathrm{mag}} = \frac{1}{\bar{m}_{\min}},
\end{equation}
with $\bar{m}_{\min}$ denoting the average signal magnitude of a tissue, minimized over tissues. Thus, $C_{\mathrm{mag}}$ penalizes pulse sequences yielding weak signals, and the strength of this penalty relative to the rest of the cost function is tuned by adjusting the coefficient $w_{\mathrm{mag}}$.

In Figure 4 of the main text and tables 1-4 of this supporting material, we present fifteen optimized pulse sequences labelled $a$ through $o$. Two of these, sequences $i$ and $j$, are produced by optimizations in which the coefficient $w_{\mathrm{mag}}$ has been set non-zero. The other thirteen optimized sequences are all produced using $w_{\mathrm{mag}} = 0$.

A stronger signal magnitude should result in a better signal to noise ratio. Thus, one may expect that the magnitude incentive is redundant with the model of random error (from \cite{R8, R9}) that is already incorporated into the calculation of $\sigma^{(T_1)}$ and $\sigma^{(T_2)}$. The $C_{\mathrm{mag}}$ term defined in \eq{eq:cmag} is not, however, manifestly equivalent to the magnitude incentive achieved indirectly through $\sigma^{(T_1)}$ and $\sigma^{(T_2)}$. Thus we chose to experimentally test some sequences optimized using such a term.

In figure 4 of the main text, one sees that in fact sequences $i$ and $j$, which were produced with nonzero $w_{\mathrm{mag}}$, achieve the best precision as measured by bootstrapping statistics applied to in vivo data. Two general classes of hypotheses might be proposed for why this is the case. One is that the incorporation of nonzero $w_{\mathrm{mag}}$ yields a cost function that achieves better modeling of the notion of precision that is measured by bootstrap statistics. The second is that nonzero $w_{\mathrm{mag}} = 0$ simply yielded better convergence of the optimizer. In other words, with $w_{\mathrm{mag}} = 0$ the optimizer converged poorly and reached solutions that were far from optimal, as defined by the cost function itself.

Cost functions like equation \ref{eq:multi}, containing a linear combination of two terms, are studied under the rubric of ``multi-objective optimization.'' As the weighting coefficient is swept, the resulting set of global optima define a Pareto-optimal tradeoff frontier. It holds rigorously that the resulting tradeoff curve must be monotonically decreasing, as any decrease in one cost term must be accompanied by an increase in the other. (If this were not the case, one could improve one term without causing any deterioration in the other, thus the point being plotted is not actually optimal.) In an optimization which is well-converged but not hitting exact global optima, this property should hold approximately.

\begin{figure}[htbp]
    \begin{center}
    \includegraphics[width=0.85\textwidth]{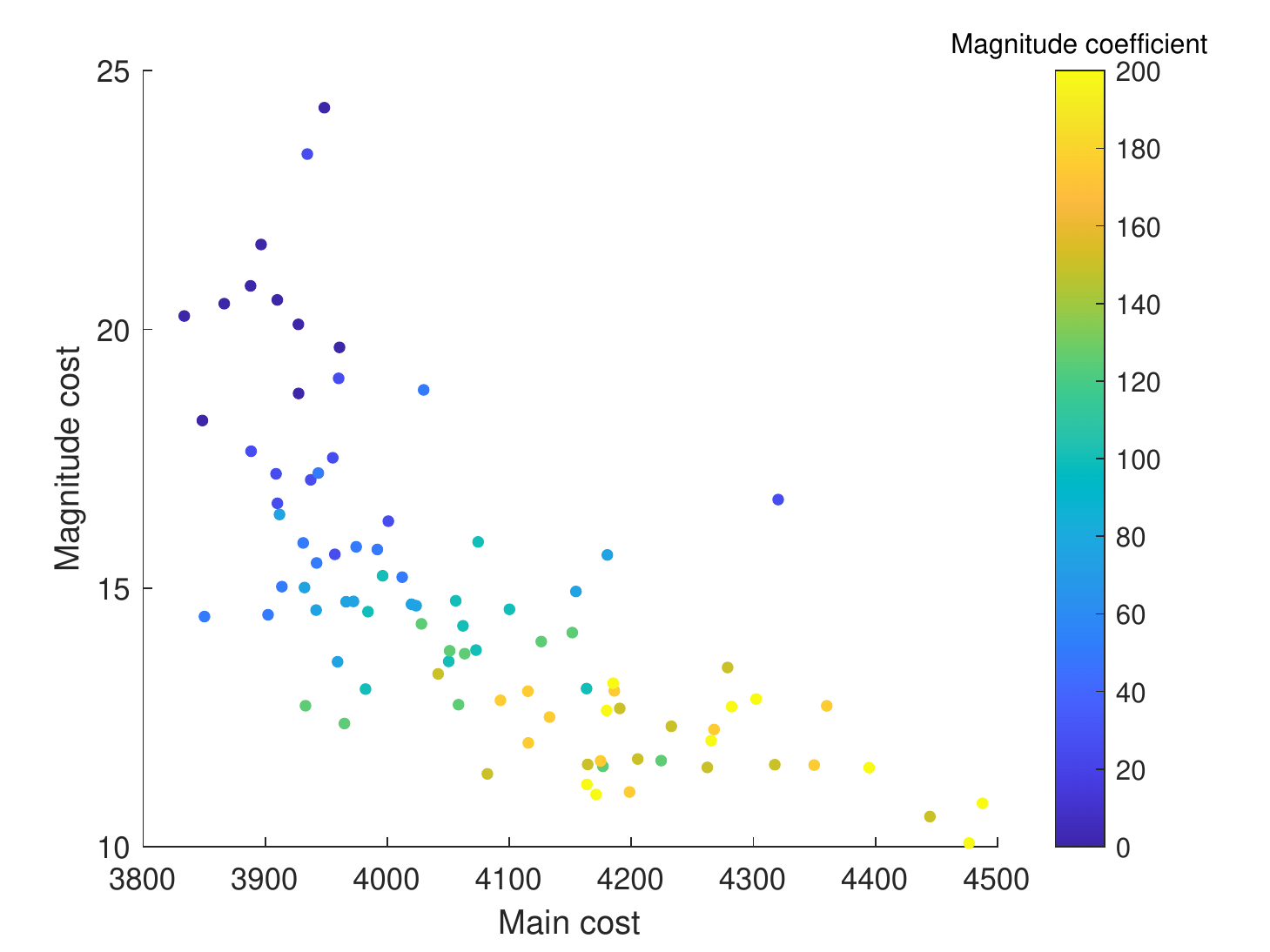}
    \end{center}
    \caption{\label{fig:mag_frontier} The cost function takes the form $C = C_{\mathrm{main}} + w_{\mathrm{mag}} C_{\mathrm{mag}}$, where $C_{\mathrm{main}}$ is the main cost to be minimized, $C_{\mathrm{mag}}$ is a term that penalizes small signal magnitude, and $w_{\mathrm{mag}}$ is a coefficient that sets the relative weight of the magnitude term. In a well-converged optimization, any decrease in the magnitude penalty will come at the cost of an increase in the main cost. By changing the magnitude coefficient one can sweep across this tradeoff curve. The general trend of the points obtained from the 90 optimization results shown here is consistent with this. However, there is a substantial amount of scatter thus illustrating that the optimization is generally not reaching exact global optima.}
\end{figure}

In figure \ref{fig:mag_frontier}, such a tradeoff frontier is shown, which displays this approximate monotonicity property, though with substantial scatter. This is consistent with the improved-modeling hypothesis but does not resolve the issue definitively. We thus leave further exploration of this issue to future work.

\section{Search Space Parameterization}
\label{sec:param}

For most of our optimizations we set $\theta_{\tri} = 0$ for all $\tri$. In this case, the number of parameters defining a pulse sequence is $2n$, where $n$ is the number of TRs. In this work we consider sequences with $480 \leq n \leq 3000$, which have duration roughly 5 seconds to 35 seconds. MRF pulse sequence design is thus a continuous-variable optimization problem on a $2n$-dimensional search space parameterized by $n$ flip angles $\alpha_1, \ldots, \alpha_n$ and $n$ durations $\mathrm{TR}_1, \ldots, \mathrm{TR}_n$. The cost function also turns out to be highly non-convex, as illustrated in figure 5 of the main text. Finding global optima for such a high-dimensional non-convex optimization problem is likely out of reach for existing algorithms and computational hardware. Furthermore, even finding good local optima is challenging on such a high-dimensional and rugged optimization landscape. 

To make a high dimensional optimization problem more manageable one can use prior knowledge to narrow the search to more promising regions of the search space. One way to do this is to initialize the optimization algorithm with a prior solution already known to be good. A different way is to use a parameterization of the search space that restricts the optimizer to explore some lower dimensional manifold of solutions thought to be promising. Here, we take this latter approach. Specifically, we restrict attention only to pulse sequences in which the flip angle and TR times vary smoothly from one TR to the next. This is motivated by the observation that such sequences typically have lower Fourier undersampling error than ``rough'' sequences \cite{R4, R7, STH17, R11}.
We achieve this by parameterizing the flip angle $\alpha$ vs. $\tri$ and $\mathrm{TR}$ vs $\tri$ curves using cubic splines\footnote{We have also tried other parameterizations: direct parameterization in terms of all $2n$ variables $(\alpha_{\tri},\mathrm{TR}_{\tri})$, piecewise linear, piecewise constant, and linear combination of Gaussians. However, all of the best sequences in terms of in vivo performance, and all of the sequences reported in this paper, come from spline parameterizations.} The spline is determined by a small number $k$ of control points (typically $10 \leq k \leq 20$) over which the optimizer has control of vertical (\emph{i.e.} $\alpha$-axis or $\mathrm{TR}$-axis) and horizontal (\emph{i.e.} $\tri$-axis) position. Because the first and last control points of each spline are pinned to $\tri = 1$ and $\tri = n$, respectively, this yields a $(2k-2)$-dimensional search space. Within the resulting search space of more manageable dimension, we generate starting points for the optimizer uniformly at random and attempt to optimize more globally. This opens the possibility of finding novel pulse sequences unbiased by any human-designed starting point.

\section{Optimization Algorithms}
\label{sec:algorithms}

\begin{figure}[htbp]
    \begin{center}
    \includegraphics[width=0.95\textwidth]{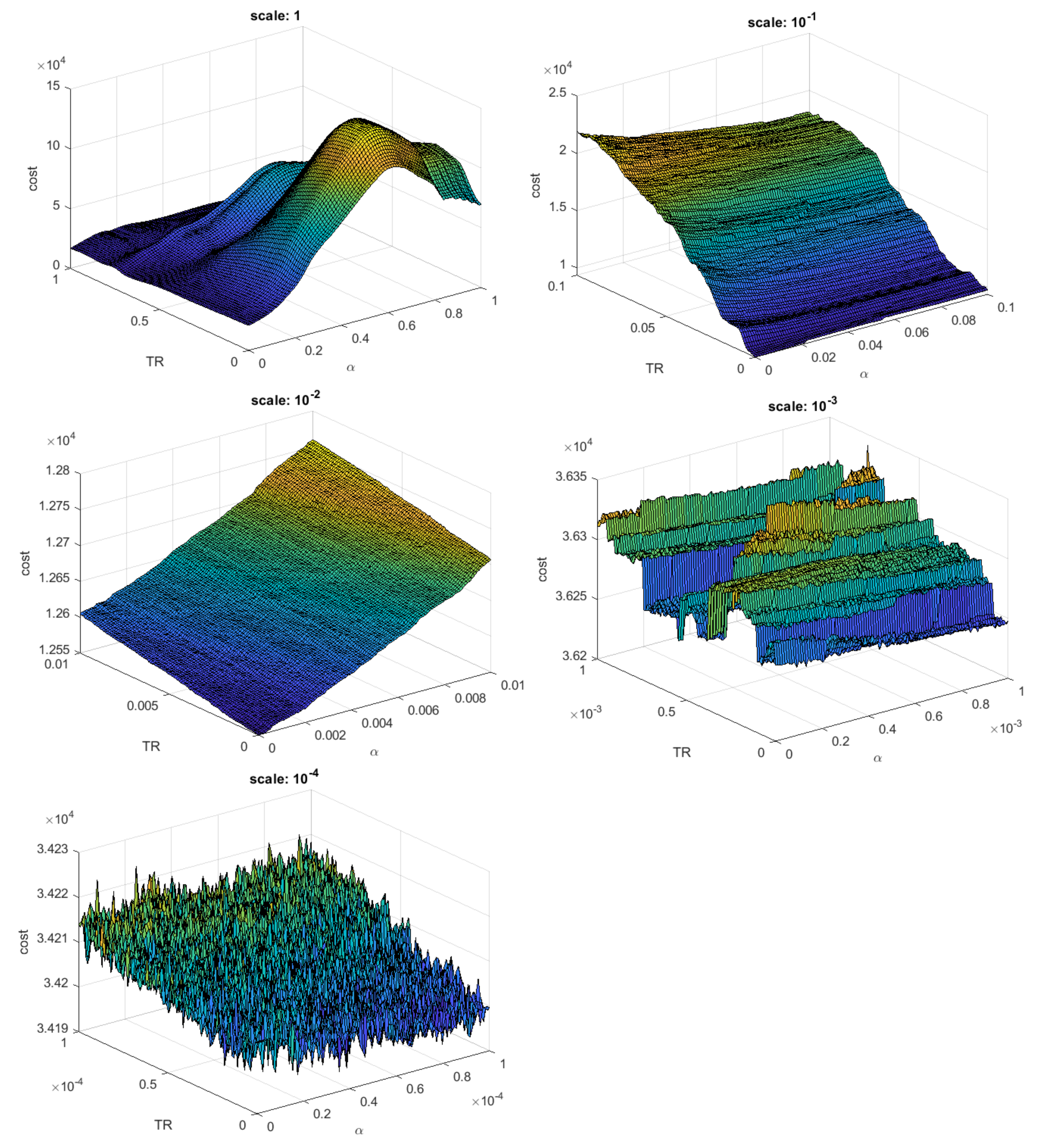}
    \end{center}
    \caption{\label{fig:landscape} Samples of the cost function landscape for a pulse sequence of 480 TRs. Here, the optimization landscape is specified by a cubic spline with 18 degrees of freedom that dictates the flip angles ($\alpha$) and a cubic spline with 18 degrees of freedom that specifies the TR times (TR). For each of the above plots, we take a random starting point in the resulting 36-dimensional space and a pair of random 18-dimensional unit vectors to determine directions of motion for the flip angle spline and TR spline. We then plot the cost function as a function of the distance moved along these two directions. The plots are zoomed in to five different scales, by factors of ten.}
\end{figure}

We formulate the design of MRF pulse sequences as a global optimization problem over continuous variables. The cost function is treated as a black box. There is no formula for the gradient of the cost function; strictly speaking, the cost function is not differentiable due to the discrete dictionary matching involved in computing Fourier undersampling errors. Due to the highly non-convex nature of the cost function, as illustrated in figure \ref{fig:landscape}, we relied on optimization heuristics capable of escaping from local minima. The best performing of these, according to our experimentation, were simulated annealing and substochastic Monte Carlo. These algorithms are simplest to formulate in the context of discrete-variable optimization problems. The special considerations needed to adapt these to the continuous-variable problem of pulse-sequence optimization are outlined in this section.

\subsection{Continuous Substochastic Monte Carlo}
\label{sec:cssmc}

Substochastic Monte Carlo is a quantum-inspired optimization method introduced in \cite{JJL16}. Substochastic Monte Carlo is inspired directly by adiabatic quantum algorithms for optimization \cite{FGG01}. In adiabatic computation, one starts with an initial Hamiltonian $H_{\textrm{init}}$, whose ground state is easy to prepare, and slowly interpolates to some final Hamiltonian $H_{\mathrm{final}}$, whose ground state encodes the solution to the computational problem at hand. When executed on ideal quantum hardware (without decoherence) the resulting dynamics is that determined by Schr\"odinger's equation
\begin{eqnarray}
\frac{d}{dt} \ket{\psi} & = & - i H(t) \ket{\psi} \\
H(t) & = & (1-s(t)) H_{\mathrm{init}} + s(t) H_{\mathrm{final}}. \label{Hgeneral}
\end{eqnarray}
Here, the function $s(t) \in [0,1]$ is the ``annealing schedule'' according to which the interpolation is performed. In the simplest case, one could proceed from $H_{\mathrm{init}}$ to $H_{\mathrm{final}}$ at a constant rate over a period of duration $T$ by using $s(t) = t/T$. Quantum adiabatic theorems \cite{JRS07, EH12} guarantee that, if the the interpolation is done sufficiently slowly, then the system will track the instantaneous ground state of $H(t)$ and thereby produce the ground state of $H_{\mathrm{final}}$, as desired. Specifically, this can be achieved with $T = O(1/\gamma^2)$, where $\gamma = \min_{0 \leq t \leq T} \gamma(t)$ and $\gamma(t)$ is the energy gap between the ground state and first excited state for $H(t)$.

Hamiltonians in which all off-diagonal matrix elements are non-positive are known as \emph{stoquastic}. By the Perron-Frobenius theorem, the ground state of any stoquastic Hamiltonian can be expressed using only real nonnegative amplitudes. Complexity-theoretic evidence suggests that adiabatic quantum computation with stoquastic Hamiltonians cannot efficiently implement universal quantum computation \cite{BDO08}. More concretely, standard folklore in the computational physics community asserts that stoquastic adiabatic processes should be efficient to simulate on classical computers using path integral or diffusion Monte Carlo methods as they do not suffer from a ``sign problem''. On the other hand, some counterexamples have been constructed in which standard path integral and diffusion Monte Carlo methods fail to converge in polynomial time when simulating such Hamiltonians \cite{H13, JJL16} and there is complexity-theoretic evidence that polynomial-time classical simulation of general stoquastic Hamiltonians is impossible \cite{H20}.

In \cite{JJL16} it was observed that, by applying a variant of diffusion Monte Carlo to simulate a quantum adiabatic optimization, one obtains a classical optimization heuristic which is competitive with state of the art solvers on a widely studied discrete optimization problem called MAXSAT. In diffusion Monte Carlo, one constructs a Markov Chain to mimic imaginary-time Schr\"odinger equation dynamics:
\begin{equation}
\label{schrodimag}
\frac{d}{dt} \ket{\psi} = -H(t) \ket{\psi}.
\end{equation}
For timestep $\delta t$ small compared to the variation of $H(t)$ one can approximately solve \eq{schrodimag} by
\begin{equation}
\label{eprod}
\ket{\psi(T)} = \prod_{j=0}^{T/\delta t} e^{-H(j\delta t) \delta t} \ket{\psi(0)},
\end{equation}
which becomes exact in the limit $\delta t \to 0$.

As $\ket{\psi(0)}$ is a vector and $e^{-H(j\delta t) \delta t} \ket{\psi(0)}$ is a matrix, \eq{eprod} looks much like a Markov chain. However, there remain two difficulties for simulating \eq{schrodimag} using a Markov Chain. The first is that $e^{-H(j\delta t) \delta t}$ has matrix elements (which in a Markov chain become transition probabilities) that are not easy to compute. The second is that $e^{-H(j\delta t) \delta t}$ is in general not a stochastic matrix, and therefore does not preserve the sum of the entries of the vector, which in a Markov chain represent probabilities that must sum to one. (Exponentially large dimension of $e^{-H(j\delta t) \delta t}$ and $\ket{\psi(T)}$ does not pose a problem for Markov Chain Monte Carlo methods because $\ket{\psi(T)}$ is not a list of numbers to be stored in memory but rather a probability distribution to be inhabited by following the specified transition probabilities.)

The solutions to these difficulties depends on the specific structure of the Hamiltonian \eq{Hgeneral}. In a continuous variable optimization problem on $n$ variables a natural choice is to take
\begin{eqnarray}
H_{\mathrm{init}} & = & -\nabla^2 \\
H_{\mathrm{final}} & = & C(x_1,\ldots,x_n)
\end{eqnarray}
where $\nabla^2$ is the Laplacian on $\mathbb{R}^n$ (\emph{i.e.} a kinetic energy term for a single particle in $n$ dimensions) and $C(x_1,\ldots,x_n)$ is a diagonal operator in the position basis (\emph{i.e.} a potential energy term). This ensures that the ground state of $H_{\mathrm{init}}$ is the uniform superposition and the ground state of $H_{\mathrm{final}}$ is a delta function centered at the minimum of $C$. In this case, using a first order Trotter-Suzuki expansion\footnote{For the discrete problems considered in \cite{JJL16} a Taylor expansion is used instead. However, this is not possible here since $\nabla^2$ is an unbounded operator.}, one obtains
\begin{equation}
\label{Trotter}
e^{-H(t) \delta t} = e^{-(1-s(t)) \nabla^2 \delta t} e^{- s(t) C \delta t} + O(\delta t^2).
\end{equation}

The operator $e^{-(1-s(t)) \nabla^2 \delta t}$ has a direct interpretation in terms of random walks. By Fourier transform on finds that, in $n$ dimensions, for any $\alpha > 0$
\begin{equation}
\bra{\vec{y}} e^{\alpha \nabla^2} \ket{\vec{x}} = \left( \frac{1}{2 \sqrt{ \alpha \pi}} \right)^n \exp \left[ -\frac{|\vec{x}-\vec{y}|^2}{4 \alpha} \right].
\end{equation}
Thus, the corresponding stochastic dynamics is to perturb the position of the random walker by a gaussian random variable of variance $2 \alpha$.

The operator $e^{- s(t) C \delta t}$ does not correspond directly to a stochastic process, since probability is not preserved. Since $C$ is diagonal in the position basis, one has
\begin{equation}
e^{-s(t) C \delta t} \ket{\vec{x}} = e^{-s(t) C(\vec{x}) \delta t} \ket{\vec{x}},
\end{equation}
where, on the lefthand side $C$ is an operator, and on the righthand side, $C(\vec{x})$ is a number, namely the cost function evaluated at $\vec{x}$. Thus, for walkers at locations with cost less than zero, the probability must grow, and the for walkers at locations with cost greater than zero the probability must shrink. As in \cite{JJL16} we implement this via birth-death dynamics. For $C(\vec{x}) > 0$ we can assign the walker at $\vec{x}$ to ``die'' with some probability and be removed from the population. For $C(\vec{x}) < 0$ we assign the walker to ``replicate'' with some probability, yielding multiple walkers at $\vec{x}$. We choose these probabilities such that the expected number of walkers at site $\vec{x}$ gets multiplied by the desired factor $e^{-s(t) C(\vec{x}) \delta t}$.

With this interpretation, \eq{Trotter} yields a prescription for an algorithm: alternate between steps where the population of walkers is perturbed according to gaussian-distributed moves, over length scales that gradually decrease over the course of the anneal, according to some schedule specified by $s(t)$, and steps where the birth-death dynamics kills off walkers at higher values of the cost function and replicates walkers at lower values of the cost function. There are however some additional subtleties to address in order to turn this into a practical algorithm. Applied to an arbitrary cost function, such a procedure generically yields a population of walkers that either collapses to zero (if the population-average value of $C(\vec{x})$ is positive) or exponentially blows up (if the population-average value of $C(\vec{x})$ is negative). One can compensate for this by replacing $C$ with $C-\langle C \rangle$, where $\langle C \rangle$ is the cost function averaged over the current distribution of walkers. Note that, in the context of Schrodinger's equation, subtracting a time-dependent constant term from the potential $C$ would result only in an unobservable global phase. Similarly, in the imaginary-time Schr\"odinger equation, such a term only affects overall magnitude of the solution vector. Thus, this adjustment does not distort the underlying physics.

In practice, this is not quite sufficient to obtain highly stable population size, so one must add a feedback loop to stabilize it. For example, we have found it effective to replace $e^{-s(t) (C-\langle C \rangle) \delta t}$ with $f e^{-s(t) (C-\langle C \rangle) \delta t}$, where
\begin{equation}
f = \left\{ \begin{array}{rl} 0.96 & \textrm{if population exceeds target} \\
                              1.05 & \textrm{otherwise}
                              \end{array} \right.
\end{equation}
The target population is then one of the hyperparameters of the optimization algorithm. In this work we have generally found it effective to set the target population at twenty walkers.

One also must choose a timestep $\delta t$. If $\delta t$ is chosen too small then in the birth-death process very few walkers will die or replicate. Hence the tendency of the dynamics toward lower values of the cost function will be very weak and walkers will move according to almost pure diffusion. If $\delta t$ is too large then almost the entire population will quickly get concentrated on the location of the walker that currently has the lowest value of the cost function. A choice of $\delta t$ in the operator $e^{-s(t) (C-\langle C \rangle) \delta t}$ which achieves a good compromise between these extremes is to take
\begin{equation}
\label{adaptive}
\delta t = \frac{C_{\max} - C_{\min}}{s(t)}.
\end{equation}
If the distribution of $C$ over the population of walkers is such the mean cost is halfway between the maximum and minimum then this ensures that the exponent $-s(t) (C-\langle C \rangle) \delta t$ lies between $-1/2$ and $1/2$. Thus the expected number of walkers on a given site will be adjusted by a factor in the range $[e^{-1/2},e^{1/2}]$. For any distribution, it is still the case that the exponent will lie between $-1$ and $1$, and thus the expected number of walkers on a given site will always be multiplied by a factor in the range $[e^{-1},e]$. (By ignoring or, through choice of $\delta t$ eliminating, the possibility of factor greater than $2$ one can simplify the algorithm slightly by eliminating the need to ever replicate a walker into more than two walkers.)

If the goal were to achieve an accurate physical simulation of a process with pre-specified anneal schedule $s(t)$, then one would use the same timestep $\delta t$ in both $e^{- s(t) C \delta t}$ and $e^{-(1-s(t)) \nabla^2 \delta t}$ in \eq{Trotter}. Here, however, our goal is instead to achieve effective optimization. Consequently, we take as user input an annealing schedule specifying the width of the gaussian update as a function of timestep. In pseudocode, one has algorithm \ref{cssmc}.

\begin{algorithm}
	\caption{Continuous-Variable Substochastic Monte Carlo \label{cssmc}}
	\begin{algorithmic}
        \STATE \textbf{REQUIRED INPUTS:}
        \STATE $[x_v^{(\min)}(\vec{x}),x_v^{(\max)}(\vec{x})]$; functions that, given $\vec{x} \in \mathbb{R}^n$, specify allowed range of $x_v$ for $v=1\ldots n$.
        \STATE $T_{\max}$; number of timesteps
        \STATE $s:\{1,\ldots,T_{\max} \} \to [0,1]$; anneal schedule
        \STATE $P_{\mathrm{target}}$; target population size
        \STATE $C:\mathbb{R}^n \to \mathbb{R}$; cost function
        \STATE \textbf{ALGORITHM:}
        \STATE Place $P_\mathrm{target}$ walkers uniformly at random in the search space
        \STATE Let $C_{\mathrm{winner}}, \vec{x}_{\mathrm{winner}}$ equal the cost and location of lowest cost walker in population
		\FOR{$t=1$ to $T_{\max}$}            
            \STATE \textbf{comment:} First, simulate $e^{-s(t) (C-\langle C \rangle) \delta t}$
            \STATE Let $C_{\min}, C_{\max}, \langle C \rangle$ equal minimum, maximum, and average cost in current population
            \FOR{$w=1$ to current population size}
                \STATE Let $\vec{x} \in \mathbb{R}^n$ be location of walker $w$
                \STATE Let $\hat{C} = (C(\vec{x}) - \langle C \rangle)/(C_{\max} - C_{\min})$
                \STATE Let $f$ equal $0.96$ if population exceeds $P_{\mathrm{target}}$, $1.05$ otherwise
                \STATE Let $q = f \times e^{-\hat{C}}$.
                \IF{$q < 1$}
                    \STATE Keep walker as-is with probability $q$, remove walker with probability $1-q$
                \ELSE
                    \IF{$q > 2$}
                        \STATE Let $q = 2$
                        \STATE Print a warning (rare in practice)
                    \ENDIF
                    \STATE \textbf{comment:} Here we know $1 \leq q \leq 2$
                    \STATE Duplicate walker with probability $q-1$, keep walker as-is with probability $2-q$
                \ENDIF
            \ENDFOR
            \STATE \textbf{comment:} Second, simulate $e^{(1-s(t)) \nabla^2 \delta t}$
            \FOR{$w=1$ to population size}
                \STATE Let $\vec{x} \in \mathbb{R}^n$ be location of walker $w$
                \FOR{$v=1$ to $n$}
                    \STATE Let $R = x_v^{(\max)}(\vec{x}) - x_v^{(\min)}(\vec{x})$ 
                    \STATE Sample $\delta$ as gaussian random variable of $\mu = 0$ and $\sigma = (1-s(t)) \times R \times 0.1$
                    \STATE Add $\delta$ to coordinate $x_v$ of walker $w$
                    \STATE Truncate coordinate $x_v$ of walker $w$ back to range $[x_v^{(\min)}(\vec{x}), x_v^{(\max)}(\vec{x})]$, if necessary
                \ENDFOR
            \ENDFOR
            \IF{lowest cost of a walker in current population is less than $C_{\mathrm{winner}}$}
                \STATE Overwrite $C_{\mathrm{winner}}$ and $\vec{x}_{\mathrm{winner}}$ with the cost and location of this walker
            \ENDIF
		\ENDFOR
        \STATE \textbf{output} $\vec{x}_{\mathrm{winner}}$
	\end{algorithmic}
\end{algorithm}

\clearpage

Some comments on algorithm \ref{cssmc}:
\begin{itemize}
\item In algorithm \ref{cssmc} we allow for the possibility that the minimum is found at any timestep. However, it is typically found in the final timestep or a near-final timestep.
\item At the initial time step we have the random walk perturb the variables by a constant fraction of the width of the search space. Here we have taken this fraction to be $0.1$. This is a somewhat arbitrary value chosen so that the initial diffusion process take jumps large enough compared to the search space to ensure good mixing of the Markov chain but small enough to ensure that truncation at the boundaries is not too common. At the final timestep the variables are perturbed by much smaller distances according to the ratio $\frac{1-s(1)}{1-s(T_{\max})}$. To optimize the variables up to some desired precision $\epsilon$ we need these final perturbations to have magnitude on the order of $\epsilon$. The ratio of the width of the search space to the desired precision $\epsilon$ is thus an important metric of the size of the search space and dictates the range over which $s$ must be swept. For our pulse sequence optimizations we typically take this ratio to be $10^4$. Minimization by exhaustive search over $n$ variables would thus require $10^{4n}$ evaluations of the cost function. With our spline parameterization $n=36$ is a typical value, and hence $10^{144}$ evaluations would be needed for exhaustive search. In our optimizations with substochastic Monte Carlo or simulated annealing we typically use on the order of $10^5$ cost function evaluations. (However, unlike exhaustive search, these algorithms are not guaranteed to find the global minimum.)
\item By examining algorithm \ref{cssmc} one can observe that, due to the adaptively chosen offset $\langle C \rangle$ and timestep $\delta t$, the dynamics of the walkers is invariant under the transformation $C(\vec{x}) \to a C(\vec{x}) + b$ for any constant $b$ and any positive constant $a$. This is very useful in practice for optimizing pulse sequences because one must typically experiment with many different cost functions which may differ widely in scale. These changing scales require no adjustment to the algorithm or to the hyperparameters. 
\end{itemize}

\subsection{Adaptive Non-Isotropic Simulated Annealing}

Simulated annealing is a widely used physics-inspired heuristic for non-convex optimization \cite{KGV83}. At each iteration a move in the search space is proposed, which is accepted with probability $\min\{1,e^{-\beta \Delta E} \}$, where $\beta$ is interpreted as an inverse temperature, and $\Delta E$ is the change in cost function, interpreted as an energy. The rate of convergence of simulated annealing algorithms depends strongly on the acceptance rate of the proposed moves. For application to simulations in statistical physics, where the goal is sampling from a Boltzmann distribution, it can be proven under certain conditions that the optimal acceptance rate is 0.234 \cite{RGG97}. For continuous variable optimization it has been conjectured based on experimental results and heuristic arguments that an acceptance rate of 0.5 may yield optimal performance \cite{VL84}.

To achieve acceptance rates near half, the proposed moves in the search space should yield changes in the value of the cost function (here interpreted as energy) of roughly comparable magnitude to the temperature $T = 1/\beta$. (For convenience we here use units where Boltzmann's constant is unity.) For typical cost functions, making smaller magnitude changes to $\vec{x}$ will yield smaller magnitude changes to $C(\vec{x})$. Consequently, efficient simulated annealing algorithms for continuous variable optimization problems propose smaller moves in the search space as the temperature is decreased \cite{VL84}.

Here, we choose step size as a function of temperature using an approach tailored to the specific properties of our pulse sequence optimization problem. As discussed in \sect{sec:param}, we have four distinct variable types: horizontal coordinates of spline control points for flip angle, vertical coordinates for spline control points for flip angle, horizontal coordinates for spline control points for TR time, and vertical coordinates for spline control points for TR time. The typical amount that the cost function is changed by making an adjustment of given magnitude to a variable can be expected to differ strongly between these four types of variables. Therefore, at the start of the optimization, for each of the four variable types, we separately perform random sampling to estimate the typical magnitude of change in cost as a function of magnitude of perturbation to those variables. 

For our specific cost function, we find that the median absolute value of change in cost, as a function of the magnitude in the change of a variable, is well fit by assuming that the magnitude of change in cost is proportional to the magnitude of the move in the search space. We thus extract four proportionality constants, one for each variable type, from random sampling and linear fits, at the start of the anneal, and then generating proposed moves by perturbing a given variable by adding a gaussian random variable with mean zero and standard deviation proportional to the temperature, via the proportionality constant appropriate to the variable's type. Precise details of our implementation of adaptive non-isotropic simulated annealing are given in algorithms \ref{perturbvar} through \ref{ANISA}. Any questions can be directed to the corresponding author \texttt{dan.ma@case.edu}.

\begin{figure}[htbp]
\begin{center}
\includegraphics[width=0.8\textwidth]{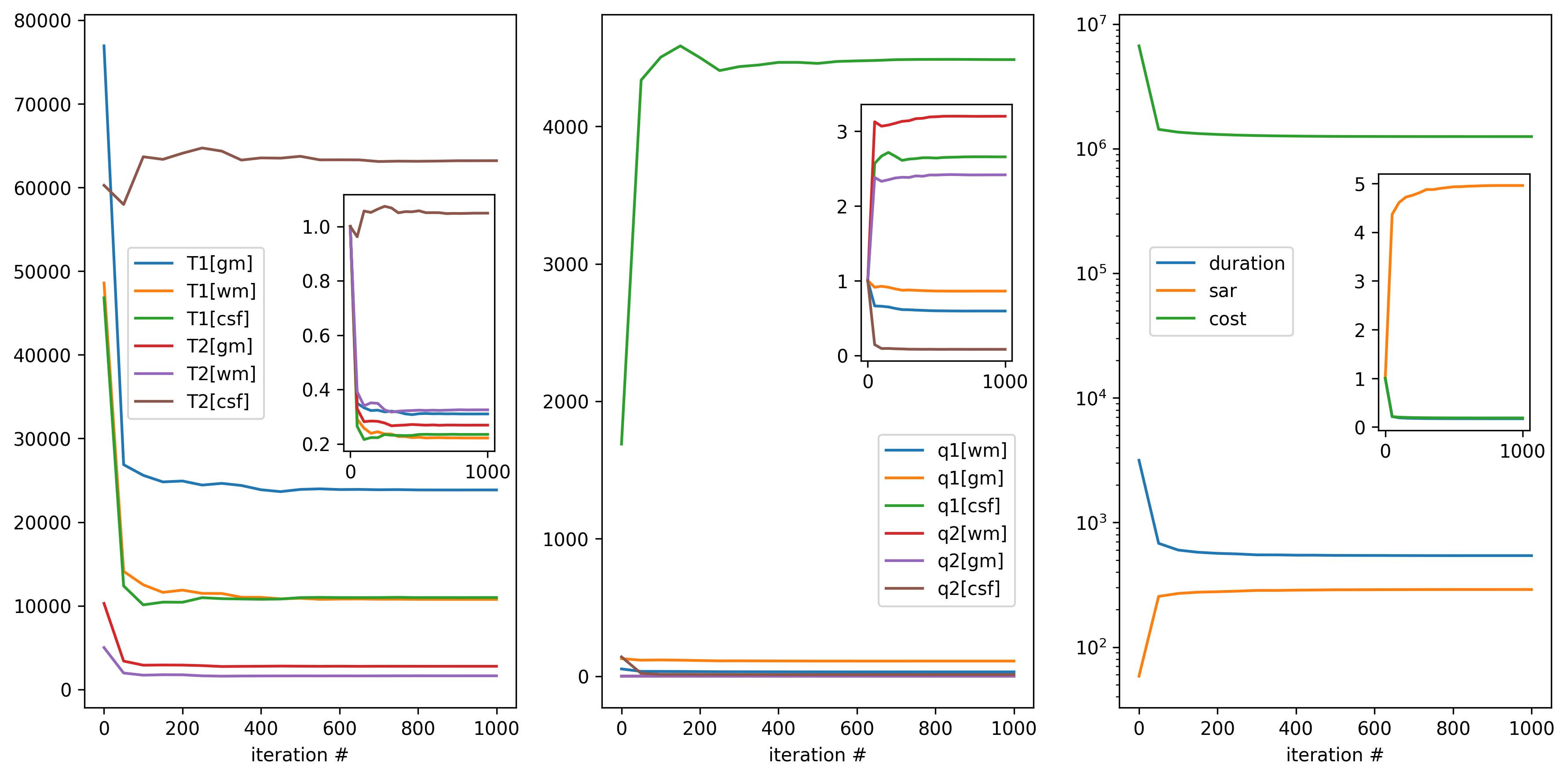}
\end{center}
\caption{\label{fig:convergence} Evolution of the three tissue $T_1$ and $T_2$ undersampling errors (a), and random errors (b) over 2000 optimization steps. All errors were normalized to the value at the first step. The biggest gains are made in the initial iterations. At later iterations small gains in cost function are obtained by moving along tradeoff curves in which one form of error is improved at the expense of another source of error whose weight coefficient in the cost function is smaller.}
\end{figure}

\begin{algorithm}
	\caption{PerturbVar$(x,x_{\min},x_{\max},\mathrm{scale})$ \label{perturbvar}}
	\begin{algorithmic}
        \STATE COMMENT: this routine modifies $x$, e.g. via pass by reference
        \STATE Let $\delta = \mathrm{Gaussian}[\mu = 0, \sigma = 1] \times \mathrm{scale} \times (x_{\max} - x_{\min})$
        \STATE Let $x = x + \delta$
        \IF{$x < x_{\min}$}
            \STATE Let $x = x_{\min}$
        \ENDIF
        \IF{$x > x_{\max}$}
        \STATE Let $x = x_{\max}$
        \ENDIF        
	\end{algorithmic}
\end{algorithm}

\begin{algorithm}
	\caption{MedianDiff$(\mathrm{scale}, \ell)$ \label{mediandiff}}
	\begin{algorithmic}
        \FOR{$t=1$ to $499$} 
            \STATE Place $\vec{x}$ uniformly at random in the search space
            \STATE Let $E_{\mathrm{before}} = C(\vec{x})$
            \STATE Choose $v$ uniformly at random among variables of type $\ell$
            \STATE PerturbVar$(\vec{x}_v, x_v^{(\min)},x_v^{(\max)}, \mathrm{scale})$
            \STATE COMMENT: $E_{\mathrm{after}}$ is evaluated at the perturbed value of $\vec{X}$. 
            \STATE Let $E_{\mathrm{after}} = C(\vec{x})$ 
            \STATE Let $\Delta_t = | E_{\mathrm{after}} - E_{\mathrm{before}} |$
        \ENDFOR
        \STATE Return $\bar{\Delta}_{\ell} = $ median of $\Delta_{\{1,\ldots,499\}}$
	\end{algorithmic}
\end{algorithm}

\begin{algorithm}
	\caption{FindScaleFactors \label{findscalefactors}}
	\begin{algorithmic}
        \FOR{$J = 1$ to $4$}
            \STATE Let $S_J = 0.03 \times e^{-2J}$
            \FOR{$\ell = 1$ to NumTypes}
                \STATE Let $\bar{\Delta}^{(\ell)}_J = \textrm{MedianDiff}(S_J, \ell)$
            \ENDFOR
        \ENDFOR
        \FOR{$\ell = 1$ to NumTypes}
            \STATE COMMENT: Least squares fit of $\bar{\Delta}^{(\ell)}_J = F_{\ell} S_J$ to $\{(\bar{\Delta}^{(\ell)}_J, S_J): J=1\ldots4\}$
            \STATE Let $F_{\ell} = \left( \sum_{J=0}^4 S_J \bar{\Delta}^{(\ell)}_J \right) \bigg/ \left( \sum_{J=0}^4 S_J^2 \right)$ 
        \ENDFOR
        \STATE Return array $F_{\ell = 1 \ldots \mathrm{NumTypes}}$
	\end{algorithmic}
\end{algorithm}

\begin{algorithm}
	\caption{Adaptive Non-Isotropic Simulated Annealing (ANISA) \label{ANISA}}
	\begin{algorithmic}
        \STATE Let $F_{\ell = 1 \ldots \mathrm{NumTypes}} = \textrm{FindScaleFactors}$
        \STATE Let $kT = 0.1 \times F_0$
        \STATE Let $r = R^{1/T_{\max}}$
        \STATE Initialize $\vec{x}$ uniformly at random in the search space
        \FOR{$t = 1$ to $T_{\max}$}
            \FOR{$\ell = 1$ to NumTypes}
                \FOR{$v$ in variables of type $\ell$}
                    \STATE PerturbVar$(\vec{x}_v, kT/F_{\ell})$
                    \STATE Accept or reject according to Metropolis rule at temperature $kT$
                \ENDFOR
            \ENDFOR
            \STATE Let $kT = kT \times r$
        \ENDFOR
	\end{algorithmic}
\end{algorithm}

\clearpage

\section{Additional Data}

The data in tables \ref{tab:raw}, \ref{tab:phant}, \ref{tab:phant_nophase}, and \ref{tab:qual} is available for download in csv format at\\
\texttt{https://github.com/madan6711/Automatic-MRF-seq-design}.

\begin{figure}[htbp]
\begin{center}
\includegraphics[width=\textwidth]{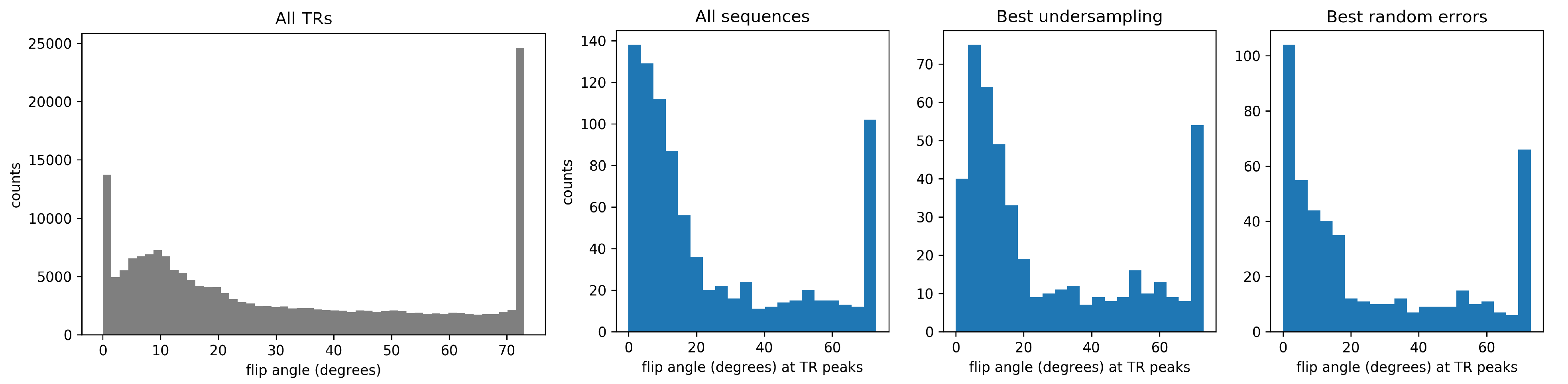}
\end{center}
\caption{\label{fig:histogram} Histograms of flip angles aggregated across an ensemble of 388 optimized pulse sequences. The maximum flip angle of 73 degrees is determined from a Sinc pulse with a duration of 2000 us and time bandwidth product of 8, which is used to limit deviation from nominal flip angles and reduce bias in the resulting maps\cite{MCC17}. Left panel shows histograms of flip angles across all TRs of all pulse sequences. Right three panels show flip angles in TRs of peak duration. One can observe that low flip angles are much more prevalent at peak duration TRs. Furthermore, this favoring of low flip angles at TR duration peaks is most pronounced in the optimized pulse sequences in which random errors were more strongly optimized at the expense of undersampling errors.}
\end{figure}

\begin{figure}[htbp]
    \begin{center}
    \includegraphics[width=\textwidth]{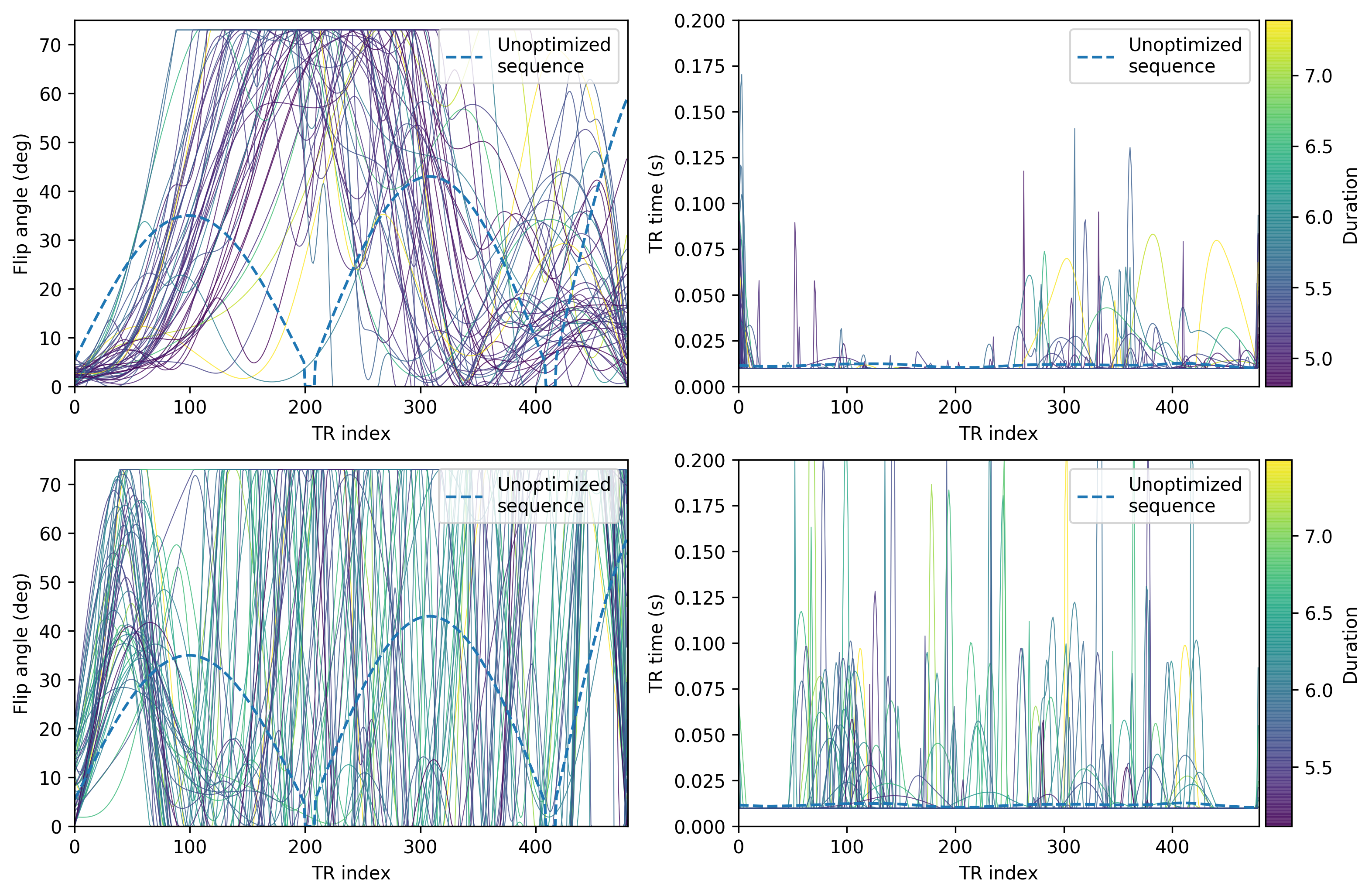} 
    \end{center}
    \caption{\label{fig:random_vs_undersampling} Flip angles and TR durations for the top 10\% optimized sequences sorted by lowest undersampling errors (top row), and by lowest random errors (bottom row). A standard unoptimized sequence (dotted curve) is added for comparison.}
\end{figure}

\begin{table}[htbp]
\begin{center}
\begin{tabular}{|l|l|l|l|l|}
\hline
Sequence                 & TRs   & Duration (s) & Avg $T_1$ error (ms) & Avg $T_2$ error (ms) \\
\hline
standard480              & 480   & 5.57         & 25.1                 & 3.51 \\  
\hline
standard672              & 672   & 7.75         & 24.6                 & 2.04 \\  
\hline
standard864              & 864   & 9.93         & 24.3                 & 1.85 \\  
\hline
standard1056             & 1056  & 12.21        & 24.1                 & 1.76 \\  
\hline
standard1248             & 1248  & 14.50        & 23.6                 & 1.64 \\  
\hline
standard1440             & 1440  & 16.69        & 23.2                 & 1.51 \\  
\hline
standard1632             & 1632  & 18.99        & 23.1                 & 1.43 \\  
\hline
standard2400             & 2400  & 28.05        & 22.3                 & 1.20 \\  
\hline
standard2592             & 2592  & 30.30        & 22.2                 & 1.17 \\  
\hline
standard2784             & 2784  & 32.52        & 21.9                 & 1.16 \\  
\hline
standard3000             & 3000  & 34.95        & 21.7                 & 1.13 \\  
\hline
optimized a              & 480   & 6.43         & 27.15                & 1.67 \\  
\hline
optimized b              & 1000  & 16.76        & 20.1                 & 1.48 \\  
\hline
optimized c              & 1000  & 15.92        & 23.7                 & 2.01 \\  
\hline
optimized d              & 480   & 11.40        & 29.7                 & 2.45 \\  
\hline
optimized e              & 480   & 10.37        & 37.4                 & 2.26 \\  
\hline
optimized f              & 480   & 14.11        & 20.3                 & 1.96 \\  
\hline
optimized g              & 480   & 10.99        & 21.1                 & 1.90 \\  
\hline
optimized h              & 480   & 8.89         & 25.1                 & 1.81 \\  
\hline
optimized i              & 480   & 7.76         & 13.2                 & 0.88 \\  
\hline
optimized j              & 960   & 13.66        & 13.2                 & 0.73 \\  
\hline
optimized k              & 480   & 5.83         & 28.1                 & 1.81 \\  
\hline
optimized l              & 480   & 5.00         & 28.7                 & 1.81 \\  
\hline
optimized m              & 480   & 5.02         & 18.9                 & 1.84 \\  
\hline
optimized n              & 480   & 4.88         & 27.1                 & 1.32 \\  
\hline
optimized o              & 480   & 5.85         & 22.1                 & 1.72 \\  
\hline
\end{tabular}
\end{center}
\caption{\label{tab:raw} Unoptimized sequences, c2p480--c2p3000, of different durations, are compared to optimized sequences. The average $T_1$ error and $T_2$ error are computed for four regions of interest in the white matter and then averaged. These are standard deviations under the influence of gaussian noise, as computed by applying the bootstrap method of \cite{RBB07} to in vivo data obtained from healthy volunteers.}
\end{table}

\begin{table}[htbp]
\begin{center}
\begin{tabular}{|l|l|l|l|l|l|l|}
\hline
sequence                 & $\sigma_1$(WM) & $\sigma_2$(WM) & $\sigma_1$(GM) & $\sigma_2$(GM) & $\sigma_1$(CSF) & $\sigma_2$(CSF) \\
\hline
standard480              & 94.4          & 47.3          & 153.7         & 69.5          & 78.3           & 873.1          \\  
\hline
standard672              & 98.3          & 12.0          & 166.1         & 27.2          & 75.8           & 451.6          \\  
\hline
standard864              & 113.8         & 8.5           & 166.7         & 26.5          & 71.8           & 381.1          \\  
\hline
standard1056             & 105.7         & 9.8           & 159.6         & 26.0          & 63.9           & 376.5          \\  
\hline
standard1248             & 90.1          & 11.7          & 148.8         & 26.7          & 57.6           & 386.4          \\  
\hline
standard1440             & 102.5         & 8.5           & 159.0         & 25.8          & 52.7           & 358.9          \\  
\hline
standard1632             & 108.1         & 8.1           & 162.5         & 24.9          & 49.5           & 363.9          \\  
\hline
standard2400             & 89.2          & 8.4           & 137.5         & 32.0          & 54.2           & 209.2          \\  
\hline
standard2592             & 79.9          & 7.5           & 128.4         & 30.8          & 51.9           & 206.7          \\  
\hline
standard2784             & 74.0          & 8.1           & 122.8         & 30.7          & 49.2           & 212.8          \\  
\hline
standard3000             & 92.9          & 7.2           & 138.3         & 30.2          & 50.4           & 233.6          \\  
\hline
optimized a              & 84.8          & 5.8           & 196.5         & 19.7          & 120.6          & 429.2          \\  
\hline
optimized b              & 84.3          & 6.3           & 139.5         & 24.2          & 97.8           & 208.8          \\  
\hline
optimized c              & 65.5          & 13.4          & 80.3          & 22.1          & 179.5          & 333.5          \\  
\hline
optimized d              & 181.5         & 5.8           & 340.3         & 63.5          & 263.4          & 341.5          \\  
\hline
optimized e              & 51.7          & 2.9           & 127.9         & 8.6           & 47.5           & 287.9          \\  
\hline
optimized f              & 47.7          & 18.1          & 84.0          & 28.1          & 60.6           & 149.9          \\  
\hline
optimized g              & 34.1          & 7.6           & 70.3          & 15.9          & 98.3           & 205.2          \\  
\hline
optimized h              & 79.2          & 6.4           & 128.2         & 21.6          & 118.8          & 416.7          \\  
\hline
optimized i              & 101.8         & 10.1          & 236.0         & 24.3          & 76.6           & 308.4          \\  
\hline
optimized j              & 322.3         & 9.3           & 376.8         & 27.0          & 74.0           & 301.6          \\  
\hline
optimized k              & 40.0          & 1.2           & 114.8         & 2.9           & 59.6           & 124.1          \\  
\hline
optimized l              & 38.1          & 1.3           & 107.4         & 2.4           & 53.0           & 176.7          \\  
\hline
optimized m              & 42.3          & 1.5           & 126.0         & 4.0           & 59.7           & 105.3          \\  
\hline
optimized n              & 26.1          & 1.9           & 81.3          & 7.0           & 136.8          & 691.0          \\  
\hline
optimized o              & 48.9          & 1.6           & 127.3         & 4.8           & 82.8           & 105.2          \\  
\hline
\end{tabular}
\end{center}
\caption{\label{tab:phant} Systematic errors as predicted by the three-tissue digital phantom. Each voxel in the simulated brain is assigned to be either white matter (WM), grey matter (GM), or cerebrospinal fluid (CSF). The bloch equations are solved, and the resulting measurement outcomes are computed as in equation \ref{3tissue} using point spread functions that incorporate phase errors. The inferred values of $T_1$ and $T_2$ are then computed from these signals for each voxel by dictionary matching. The root-mean-square deviation of the inferred value from the original value assigned to the voxels is tabulated for $T_1$ for $T_2$ for each of the three tissue types. $\sigma_j(T)$ is the RMS deviation in $T_j$ for tissue type $T$, expressed in milliseconds.}
\end{table}

\begin{table}[htbp]
\begin{center}
\begin{tabular}{|l|l|l|l|l|l|l|}
\hline
sequence                 & $\sigma_1$(WM) & $\sigma_2$(WM) & $\sigma_1$(GM) & $\sigma_2$(GM) & $\sigma_1$(CSF) & $\sigma_2$(CSF) \\
\hline
standard480              & 57.3           & 7.3            & 148.1          & 18.7           & 77.5            & 615.6           \\  
\hline
standard672              & 54.3           & 4.3            & 136.7          & 26.3           & 79.6            & 337.6           \\  
\hline
standard864              & 54.0           & 4.4            & 131.5          & 27.1           & 74.5            & 284.0           \\  
\hline
standard1056             & 53.4           & 4.2            & 130.5          & 24.7           & 64.7            & 289.5           \\  
\hline
standard1248             & 51.2           & 4.4            & 127.9          & 24.9           & 58.7            & 306.0           \\  
\hline
standard1440             & 50.7           & 4.5            & 125.8          & 26.2           & 54.8            & 285.0           \\  
\hline
standard1632             & 50.1           & 4.5            & 123.4          & 25.6           & 50.8            & 277.4           \\  
\hline
standard2400             & 45.4           & 5.2            & 98.7           & 35.7           & 55.5            & 181.9           \\  
\hline
standard2592             & 43.8           & 5.2            & 96.6           & 34.5           & 52.1            & 192.8           \\  
\hline
standard2784             & 42.8           & 5.1            & 95.5           & 34.1           & 48.8            & 204.3           \\  
\hline
standard3000             & 43.0           & 4.9            & 93.7           & 33.5           & 47.4            & 208.4           \\  
\hline
optimized a              & 31.6           & 2.9            & 99.4           & 22.1           & 83.6            & 319.8           \\  
\hline
optimized b              & 42.5           & 4.5            & 78.1           & 27.0           & 92.6            & 127.2           \\  
\hline
optimized c              & 35.7           & 4.7            & 67.8           & 21.1           & 87.3            & 284.2           \\  
\hline
optimized d              & 49.8           & 1.6            & 116.0          & 6.0            & 131.7           & 60.2            \\  
\hline
optimized e              & 42.4           & 1.5            & 119.0          & 2.5            & 46.9            & 44.1            \\  
\hline
optimized f              & 32.3           & 4.6            & 58.0           & 23.7           & 49.5            & 136.0           \\  
\hline
optimized g              & 31.7           & 4.2            & 59.5           & 17.3           & 19.0            & 187.7           \\  
\hline
optimized h              & 32.4           & 5.0            & 60.2           & 23.5           & 38.4            & 281.5           \\  
\hline
optimized i              & 42.4           & 5.0            & 95.5           & 25.5           & 52.4            & 281.8           \\  
\hline
optimized j              & 48.6           & 8.1            & 110.5          & 33.5           & 68.6            & 207.6           \\  
\hline
optimized k              & 42.3           & 1.4            & 119.9          & 3.3            & 53.8            & 199.1           \\  
\hline
optimized l              & 42.2           & 1.4            & 111.0          & 2.3            & 46.8            & 328.4           \\  
\hline
optimized m              & 44.1           & 1.5            & 133.5          & 4.2            & 55.7            & 148.6           \\  
\hline
optimized n              & 29.0           & 1.9            & 81.1           & 7.8            & 130.3           & 781.2           \\  
\hline
optimized o              & 53.5           & 1.5            & 135.7          & 5.4            & 73.4            & 311.1           \\  
\hline
\end{tabular}
\end{center}
\caption{\label{tab:phant_nophase} Here, digital phantom predictions are tabulated as in table \ref{tab:phant} except that this model assumes no phase errors.}
\end{table}

\begin{table}[htbp]
\begin{center}
\begin{tabular}{|l|l|l|l|l|l|l|l|}
\hline
sequence             & min mag  & $q_1$(WM) & $q_2$(WM)   & $q_1$(GM)   & $q_2$(GM)   & $q_1$(CSF)  & $q_2$(CSF)  \\
\hline
standard480          & 0.0587 & 3.90e-01 & 2.73e-04 & 1.49e+00 & 6.05e-04 & 3.43e+01 & 3.83e-02 \\  
\hline
standard672          & 0.0547 & 3.99e-01 & 3.92e-04 & 1.53e+00 & 9.03e-04 & 4.69e+01 & 2.60e-01 \\  
\hline
standard864          & 0.0547 & 4.04e-01 & 4.81e-04 & 1.55e+00 & 1.05e-03 & 4.72e+01 & 3.28e-01 \\  
\hline
standard1056         & 0.0570 & 4.09e-01 & 5.31e-04 & 1.55e+00 & 1.12e-03 & 4.90e+01 & 3.30e-01 \\  
\hline
standard1248         & 0.0588 & 4.25e-01 & 6.46e-04 & 1.57e+00 & 1.37e-03 & 5.47e+01 & 3.54e-01 \\  
\hline
standard1440         & 0.0594 & 4.34e-01 & 7.63e-04 & 1.59e+00 & 1.58e-03 & 5.71e+01 & 4.11e-01 \\  
\hline
standard1632         & 0.0600 & 4.40e-01 & 8.66e-04 & 1.60e+00 & 1.75e-03 & 5.82e+01 & 4.31e-01 \\  
\hline
standard2400         & 0.0575 & 4.76e-01 & 1.36e-03 & 1.68e+00 & 2.67e-03 & 6.34e+01 & 2.12e+00 \\  
\hline
standard2592         & 0.0583 & 4.88e-01 & 1.42e-03 & 1.69e+00 & 2.78e-03 & 6.39e+01 & 2.17e+00 \\  
\hline
standard2784         & 0.0591 & 4.98e-01 & 1.48e-03 & 1.71e+00 & 2.87e-03 & 6.53e+01 & 2.30e+00 \\  
\hline
standard3000         & 0.0596 & 5.06e-01 & 1.57e-03 & 1.72e+00 & 3.04e-03 & 6.75e+01 & 2.47e+00 \\  
\hline
optimized a          & 0.0617 & 2.24e-01 & 5.12e-04 & 7.40e-01 & 1.01e-03 & 1.74e+01 & 2.31e-01 \\  
\hline
optimized b          & 0.0593 & 7.04e-01 & 9.16e-04 & 2.26e+00 & 2.16e-03 & 3.35e+01 & 1.52e+00 \\  
\hline
optimized c          & 0.0629 & 6.82e-01 & 1.21e-03 & 1.98e+00 & 3.03e-03 & 2.53e+01 & 6.61e-01 \\  
\hline
optimized d          & 0.0506 & 2.61e-01 & 2.66e-04 & 1.16e+00 & 2.82e-04 & 9.78e+00 & 1.82e-01 \\  
\hline
optimized e          & 0.0564 & 1.68e-01 & 3.58e-04 & 1.16e+00 & 4.63e-04 & 4.27e+01 & 2.73e-02 \\  
\hline
optimized f          & 0.0774 & 5.96e-01 & 8.60e-04 & 1.59e+00 & 2.37e-03 & 2.89e+01 & 2.79e+00 \\  
\hline
optimized g          & 0.0678 & 4.83e-01 & 7.94e-04 & 1.27e+00 & 1.92e-03 & 2.22e+01 & 7.85e-01 \\  
\hline
optimized h          & 0.0666 & 4.20e-01 & 8.06e-04 & 1.23e+00 & 1.93e-03 & 1.49e+01 & 4.78e-01 \\  
\hline
optimized i          & 0.0681 & 4.05e-01 & 8.25e-04 & 1.22e+00 & 2.03e-03 & 3.76e+01 & 7.38e-01 \\  
\hline
optimized j          & 0.0592 & 4.70e-01 & 1.40e-03 & 1.42e+00 & 3.24e-03 & 5.79e+01 & 1.89e+00 \\  
\hline
optimized k          & 0.0415 & 1.56e-01 & 2.97e-04 & 1.20e+00 & 3.47e-04 & 2.35e+01 & 1.19e-02 \\  
\hline
optimized l          & 0.0324 & 1.36e-01 & 2.77e-04 & 1.07e+00 & 3.45e-04 & 1.91e+01 & 6.17e-03 \\  
\hline
optimized m          & 0.0406 & 2.85e-01 & 2.88e-04 & 1.43e+00 & 3.80e-04 & 1.67e+01 & 1.85e-02 \\  
\hline
optimized n          & 0.0376 & 7.89e-02 & 3.67e-04 & 2.93e-01 & 6.18e-04 & 7.09e+00 & 6.55e-03 \\  
\hline
optimized o          & 0.0500 & 2.04e-01 & 3.22e-04 & 1.35e+00 & 3.88e-04 & 2.78e+01 & 1.65e-02 \\  
\hline
\end{tabular}
\end{center}
\caption{\label{tab:qual} For each of the three tissue types, the magnitude of the magnetization, as predicted by the Bloch equations, is averaged over the measurements (of which there is one for each TR). The minimum of these three numbers is recorded as ``min mag''. The quality factors for the three tissues are metrics of robustness against random error, which are estimated by a first order perturbative calculation in \cite{R8, R9}. The noise model is identical independently distributed complex gaussian noise of mean zero and standard deviation $\sigma_\eta$ added to the data point associated with each point in $k$-space, at each measurement. In this approximation, the predicted standard deviation in the value of $T_1$ for grey matter due to random noise is given by $\sigma_\eta/\sqrt{q_1(GM)}$, and similarly for $T_2$ and for the other tissue types.}
\end{table}

\begin{figure}[htbp]
\begin{center}
\includegraphics[width=2in]{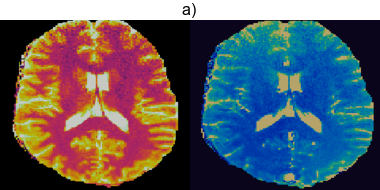}\includegraphics[width=2in]{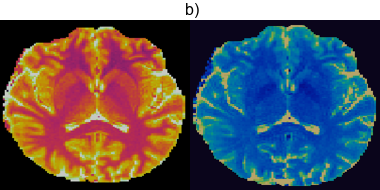}\includegraphics[width=2in]{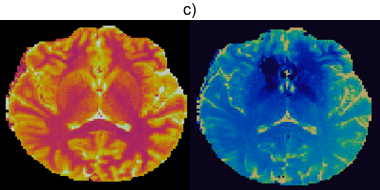}
\includegraphics[width=2in]{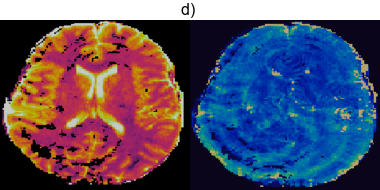}\includegraphics[width=2in]{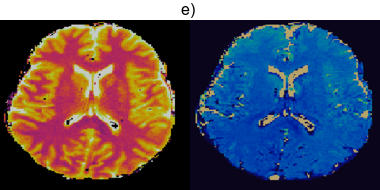}\includegraphics[width=2in]{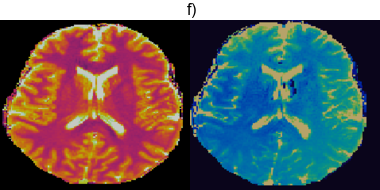}
\includegraphics[width=2in]{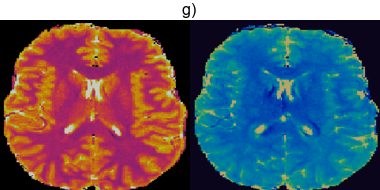}\includegraphics[width=2in]{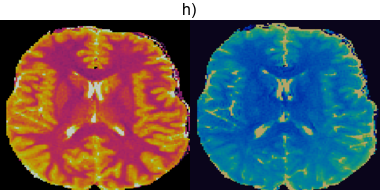}\includegraphics[width=2in]{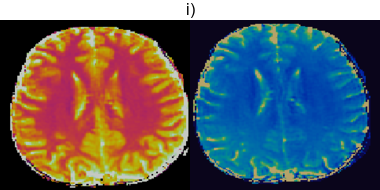}
\includegraphics[width=2in]{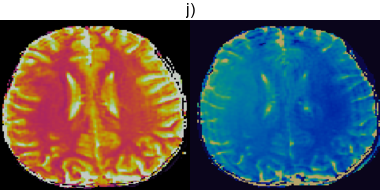}\includegraphics[width=2in]{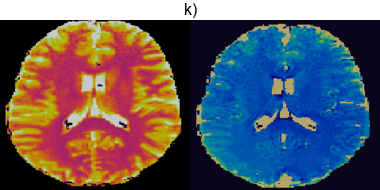}\includegraphics[width=2in]{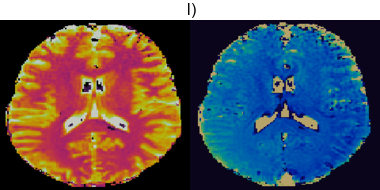}
\includegraphics[width=2in]{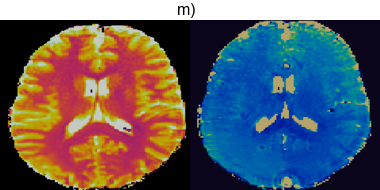}\includegraphics[width=2in]{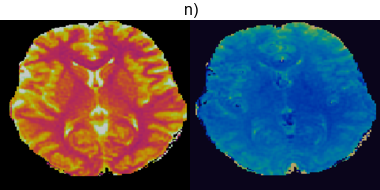}\includegraphics[width=2in]{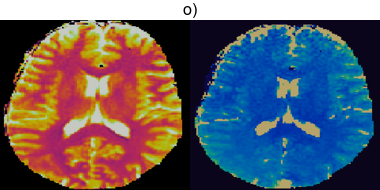}
\end{center}
\caption{\label{fig:in-vivo-images} $T_1$ (red) and $T_2$ (blue) in vivo image pairs for each of the 15 optimized sequences. The labels match the data points in figure 3 of the main text. The sequence \emph{d} was optimized with a direct parameterization rather than splines. This results in flip angles that vary less smoothly from one pulse to the next, and thereby produces stronger Fourier undersampling artifacts and inferior image quality relative the the other sequences, which are optimized using spline parameterizations.}
\end{figure}

\end{appendices}

\clearpage

\bibliography{mrf_qio}


\end{document}